\documentclass[a4paper,11pt]{article}
\pdfoutput=1

\usepackage{jheppub}
\usepackage[T1]{fontenc}
\usepackage{changes,color}
\usepackage{textgreek}

\title{\boldmath Low-reheating scenario in dark Higgs inflation and its impact on dark photon dark matter production}

\author[a,b]{Sarif Khan,}
\author[c,d,*]{Jinsu Kim,}
\note[*]{Corresponding author.}
\author[e]{and Pyungwon Ko}

\affiliation[a]{
	Department of Physics, Aliah University,\\
	Kolkata 700160, India
}
\affiliation[b]{
	Department of Physics, Chung-Ang University,\\
	Seoul 06974, Korea
}
\affiliation[c]{
    School of Physics, Faculty of Basic Sciences, University of Shanghai for Science and Technology,\\
    Shanghai 200093, China
}
\affiliation[d]{
	School of Physics Science and Engineering, Tongji University,\\
	Shanghai 200092, China
}
\affiliation[e]{
	School of Physics, Korea Institute for Advanced Study,\\
	85 Hoegi-ro, Seoul 02455, Republic of Korea
}

\emailAdd{sarifkhan@cau.ac.kr}
\emailAdd{kimjinsu@usst.edu.cn}
\emailAdd{pko@kias.re.kr}

\abstract{
We investigate dark matter (DM) phenomenology and cosmic inflation within a unified framework based on a dark $U(1)_D$ gauge extension of the Standard Model (SM). The associated dark gauge boson, namely the dark photon, serves as a viable DM candidate, which we call dark photon dark matter (DPDM), whilst the dark Higgs field drives inflation. We explore a low-reheating scenario where DM production occurs during reheating, resulting in significant entropy dilution of the DPDM abundance. Both weakly interacting massive particle (WIMP) and feebly interacting massive particle (FIMP) DM scenarios are explored, depending on the dark gauge coupling strength. For FIMP-type DM, the entropy dilution allows for stronger couplings whilst maintaining the correct relic abundance, potentially bringing these candidates within the reach of current and near-future detection experiments. Similarly, WIMP-type DM can be realised with weaker couplings. We perform a comprehensive parameter scan incorporating constraints from collider data, DM direct and indirect detection experiments, and cosmological observations. Taking quantum corrections and running of the couplings into account, we demonstrate that dark Higgs inflation yields predictions for the spectral index $n_s$ and the tensor-to-scalar ratio $r$ that are consistent with the Planck, BICEP/Keck, and ACT data. The nonminimal coupling of the dark Higgs inflaton field to gravity is shown to be much smaller than in the case of the SM Higgs inflation scenario, avoiding unitarity concerns. We show that reheating temperatures as low as 1 GeV and 1 MeV can be achieved through the decay and scattering processes of the inflaton, respectively, with the latter allowing for larger Higgs mixing angles and enhanced detection prospects. Our results establish that this minimal extension successfully unifies DM physics with inflationary cosmology.
}

\begin{document}
\maketitle
\flushbottom

\section{Introduction}
\label{sec:intro}
It is well known to us that the Standard Model (SM) is a very successful theory, with a few important drawbacks, such as the absence of a suitable dark matter (DM) candidate and neutrino masses and mixings. The presence of nonbaryonic DM is confirmed by many independent pieces of evidence, which are mostly gravitational in nature, ranging from the flatness of galaxy rotation curves \cite{Sofue:2000jx} and the Bullet Cluster \cite{Clowe:2003tk} to the precision measurements of DM abundance from the Planck data \cite{Planck:2018vyg}. There have been many beyond-the-SM (BSM) explorations aimed at tackling DM physics with very rich detection prospects at various terrestrial and space-based detectors over the last few decades \cite{Arcadi:2017kky}. The most popular scenario in this direction is the weakly interacting massive particle (WIMP) scenario, which has been searched for in all types of experiments, from colliders \cite{ATLAS:2016bek, ATLAS:2015ciy, Choubey:2017yyn, Belanger:2021slj, Biswas:2021dan, Covi:2025erx} to direct \cite{XENON:2018voc, PandaX-4T:2021bab, LZ:2024zvo} and indirect detections \cite{Fermi-LAT:2011vow, HESS:2012oad, Fermi-LAT:2015qzw, HESS:2016glm, MAGIC:2011nta}. Unfortunately, so far, no direct clues have been obtained regarding the particle nature of DM due to null detections. The leading experiment in DM searches is the direct detection experiment, which has ruled out DM interactions with the visible sector at the weak scale \cite{XENON:2018voc, PandaX-4T:2021bab, LZ:2024zvo}. However, these experiments have ruled out only particular and simple scenarios amongst many; for example, multi-component DM \cite{Costa:2022oaa, Covi:2022hqb, Belanger:2022gqc, Costa:2022lpy, Khan:2023uii, Choubey:2024krp, Khan:2024biq, Khan:2025yko, Covi:2025erx} or DM dominantly annihilating into hidden sectors \cite{Khan:2023uii, Khan:2024biq, Khan:2025yko, Covi:2025erx, Khan:2025sdl} are fully safe from all bounds. Recently, another alternative scenario, namely the feebly interacting massive particle (FIMP) scenario, has become very popular. The FIMP scenario has reduced detection prospects due to its extremely weak interactions with the visible sector \cite{McDonald:2001vt, Asaka:2005cn, Hall:2009bx}, although it still has detection prospects in low-reheating scenarios \cite{Bhattiprolu:2022sdd, Cosme:2023xpa, Silva-Malpartida:2023yks, Haque:2023yra, Arcadi:2024wwg, Lee:2024wes, Boddy:2024vgt, Haque:2024zdq, Arias:2025nub, Khan:2025keb, Bernal:2025qkj, Mondal:2025awq, Arias:2025tvd}.

Cosmic inflation, which provides not only a natural and economical explanation for the standard Hot Big Bang problems, including the flatness problem, the horizon problem, and the monopole problem, but also initial conditions for the large-scale structures of our Universe \cite{Starobinsky:1980te, Guth:1980zm, Linde:1981mu, Albrecht:1982wi, Linde:1983gd}, is also another important ingredient in explaining our Universe. The period of rapid expansion can be achieved when a scalar field, dubbed the inflaton, is dominated by its scalar potential, undergoing a phase of slow roll in the early Universe, making the Hubble parameter approximately constant and causing the scale factor to increase significantly. The SM is equipped with one scalar field, the SM Higgs field, and thus, studies using the SM Higgs field as the inflaton have naturally emerged. The quartic scalar potential, which is a good approximation of the SM Higgs potential during the large-field inflation regime, is, however, already ruled out by the cosmic microwave background (CMB) experiments, as the tensor-to-scalar ratio $r$ turns out to be too large to be compatible with the observational data \cite{Planck:2018jri, BICEP:2021xfz, ACT:2025tim}. On the other hand, if the inflaton couples to gravity through the so-called nonminimal coupling, the tensor-to-scalar ratio $r$ becomes suppressed, and the SM Higgs inflation model may become compatible with the observational data \cite{Futamase:1987ua, Fakir:1990eg, Cervantes-Cota:1995ehs, Komatsu:1999mt, Bezrukov:2007ep, Park:2008hz, Rubio:2018ogq, Cheong:2021vdb}.

The seemingly disconnected physics of DM, which is low-energy scale physics, and inflation, which is high-energy scale physics, could be linked together through the renormalisation group equations (RGEs). As such, investigations into how they are connected and studies on the effects of one on the other have been carried out by many researchers; see, for instance, Refs.~\cite{Lerner:2009xg, Gong:2012ri, Haba:2014zda, Kim:2014kok, Aravind:2015xst, Choudhury:2015eua, Tenkanen:2016idg, Ballesteros:2016xej, Hooper:2018buz, Choi:2019osi, Choi:2020ara, Kawai:2021tzc, Ghoshal:2022jeo, Cheng:2022hcm, Qi:2023egb, Kawai:2023vjo, Khan:2023uii}. Naturally, as both the DM experimental bounds and the inflation-related observational bounds, as well as theoretical constraints, are at play simultaneously, the available parameter space is heavily constrained.

In the present work, we study the DM phenomenology and inflation within a single, unified framework. For inflation, we consider the scenario where the inflaton nonminimally couples to the Ricci scalar. For the DM production mechanism, we consider both the WIMP and FIMP, depending on the interaction strength. Furthermore, we discuss DM production during reheating and the realisation of low reheating temperatures, which arise due to the delayed decay of the inflaton into the SM bath. One of the effects of a low reheating temperature is entropy production, which dilutes the DM abundance if it is present during reheating. Therefore, the crucial effect of a low reheating temperature is that the DM can be produced by the freeze-in mechanism with a stronger coupling, and similarly, for the freeze-out mechanism, we can lower the coupling strength. In the standard case with a high reheating temperature, the overproduction of DM becomes problematic. However, because of the low reheating temperature consideration, entropy production during reheating ensures the dilution of the DM abundance, bringing it within the allowed range. Thus, we may have freeze-in DM with detection prospects in ongoing experiments. On the other hand, when DM is produced via the freeze-out mechanism, choosing a weaker interaction strength would overproduce the WIMP DM in the standard scenario. With a low reheating temperature again, however, the abundance of DM is diluted through entropy production. It should be noted that the present scenario is different from the other types of DM production with a low reheating temperature \cite{Bhattiprolu:2022sdd, Cosme:2023xpa, Silva-Malpartida:2023yks, Arcadi:2024wwg, Lee:2024wes, Arias:2025nub, Khan:2025keb}, where the DM starts evolving from the low reheating temperature, and the abundance of DM is controlled mainly by the suppression of the number density due to the large mass of the relevant particles compared to the temperature.

In this perspective, we consider a popular dark $U(1)_D$ gauge extension of the SM. The $U(1)_D$-associated gauge boson, namely the dark photon, becomes the DM candidate, which we call dark photon DM (DPDM). Within the same framework, we obtain both the WIMP- and FIMP-type DM in low-reheating scenarios, depending on the value of the gauge coupling and the reheating temperature. The same setup can also accommodate a successful period of inflation. We assume that the dark $U(1)_D$ Higgs field plays the role of the inflaton and explore the inflation-related observables, such as the scalar spectral index $n_s$ and the tensor-to-scalar ratio $r$. These predictions are compared with the latest observational bounds, including the Planck, BICEP/Keck, and ACT experimental data. In addition, we discuss the realisation of a low reheating temperature in this dark Higgs inflation scenario. Through the RGEs, the running of the model parameters is taken into account to link the inflationary physics and the DM phenomenology.

The rest of the paper is organised as follows: In section~\ref{sec:model}, we outline the model, setting notations and identifying the DM candidate. The DM physics is then discussed in section~\ref{sec:dm}. Section~\ref{sec:inflation} discusses the scenario where inflation takes place along the dark $U(1)_D$ Higgs field direction. In section~\ref{sec:results}, we perform a numerical scan over a wide range of the model parameters and present the results on the viable parameter space that successfully explains both DM physics and inflationary physics. We conclude in section~\ref{sec:conc}.

\section{Model}
\label{sec:model}
We consider a dark $U(1)_D$ extension of the SM which is described by the following Lagrangian:
\begin{align}
	\mathcal{L} &= 
	\mathcal{L}_{\overline{\rm SM}}
    -\frac{1}{4} F^{D}_{\mu\nu} F^{D\mu\nu}
    -\frac{\epsilon}{2} F^{D}_{\mu\nu} F^{Y\mu\nu}
	-|D\phi_D|^2
	-|D H|^2
	-V(\phi_D,H)
	\,,
\end{align}
where $\mathcal{L}_{\overline{\rm SM}}$ represents the SM Lagrangian excluding the SM Higgs sector, the second term corresponds to the kinetic term for the dark gauge boson $W_D$, the third term describes the mixing between the dark gauge field strength $F^D_{\mu\nu}$ and the hypercharge field strength tensor $F^Y_{\mu\nu}$ with the mixing angle $\epsilon$, and $H$ ($\phi_D$) is the SM (dark) Higgs field. Here, for the dark Higgs field, the covariant derivative is defined as $D_{\mu}\phi_D = \partial_{\mu} \phi_D - i g_{D} W_{D\mu} n_{\phi_D} \phi_D$, where $g_{D}$ is the dark $U(1)_D$ gauge coupling, and $n_{\phi_D}$ is the $U(1)_D$ charge of the dark Higgs field. Without loss of generality, we choose $n_{\phi_D} = 1$.
The potential takes
\begin{align}
	V(\phi_D,H) = 
	- \mu^2_D \phi^{\dagger}_D \phi_D
	+ \lambda_D (\phi^{\dagger}_D \phi_D)^2 
	- \mu^2_H H^{\dagger} H
	+ \lambda_H (H^{\dagger} H)^2 
	+ \lambda_{HD} \phi^{\dagger}_D \phi_D H^{\dagger} H
	\,.\label{eqn:scalar_pot_full}
\end{align}
Working in unitary gauge, the dark Higgs field and the SM Higgs field can be expressed as
\begin{align}
	H = \frac{1}{\sqrt{2}}\begin{pmatrix}
		0\\
		v_{h} + h
	\end{pmatrix}
	\,,\qquad
	\phi_D = \frac{v_{D} + \phi}{\sqrt{2}}
	\,,
\end{align}
where $v_h$ and $v_D$ are the vacuum expectation values (VEVs) of the SM Higgs field and the dark $U(1)_D$ Higgs field, respectively. The charged scalar and CP-odd components become the longitudinal components of the $W, Z$ bosons and the additional $U(1)_D$ gauge boson $W_D$. In the basis of $(h\;\phi)^{T}$, the Higgs mass matrix is given by
\begin{align}
	M_{h\phi}^2 = \begin{pmatrix}
		2 \lambda_{H} v^2_{h} & \lambda_{HD} v_{h} v_{D} \\
		\lambda_{HD} v_{h} v_{D} & 2 \lambda_{D} v^2_{D}
	\end{pmatrix}
	\,.
\end{align}
We can diagonalise the above mass matrix and define a new mass basis $(h_{1}\;h_{2})^{T}$ which can be written in terms of the flavour eigenbasis $(h\;\phi)^{T}$ as follows:
\begin{align}
	\begin{pmatrix}
		h_{1}\\
		h_{2}
	\end{pmatrix} = 
	\begin{pmatrix}
		\cos\alpha & -\sin\alpha \\
		\sin\alpha & \cos\alpha
	\end{pmatrix}
\begin{pmatrix}
		h\\
		\phi
	\end{pmatrix}
    \,,
\end{align}
where the Higgs mixing angle $\alpha$ can be expressed as
\begin{align}
	\tan 2\alpha = \frac{ \lambda_{HD} v_{h} v_{D} }{\lambda_{D} v^2_{D} - \lambda_{H} v^2_{h} }\,.
\end{align}
The quartic couplings $\lambda_{H}$, $\lambda_{D}$, and $\lambda_{HD}$ can be written in terms of the physical scalar masses, $M_{h_1}$ and $M_{h_2}$, and the mixing angle, $\alpha$, as
\begin{align}
	\lambda_{H} &= \frac{M^2_{h_{1}} + M^2_{h_{2}} + \left(M^2_{h_{1}} - M^2_{h_{2}} \right)\cos2\alpha }{4 v^2_{h}}
	\,,\nonumber\\
	\lambda_{D} &= \frac{M^2_{h_{1}} + M^2_{h_{2}} - \left(M^2_{h_{1}} - M^2_{h_{2}} \right)\cos2\alpha }{4 v^2_{D}}
	\,,\nonumber\\
	\lambda_{HD} &= \frac{ \left( M^2_{h_2} - M^2_{h_1} \right) \cos\alpha \sin\alpha }{v_{h} v_{D}}
	\,.
\end{align}
We assume that the dark Higgs mass, $M_{h_2}$, is larger than the SM Higgs mass, $M_{h_1} \simeq 125$ GeV. As such, positive quartic couplings are considered in the current work.
The dark gauge boson $W_D$, whose mass is given by $M_{W_D} = g_{D} v_{D}$, becomes the DM candidate. In stabilising the DM, we use the charge conjugation symmetry in the dark sector particles which transforms the scalar and the vector particles in the dark sector as
\begin{align}
	\phi_{D} \rightarrow \phi^{\dagger}_{D}
	\,,\qquad
	W_{D\mu} \rightarrow - W_{D\mu}
	\,.
\end{align}
Alternatively, one may assume that the kinetic mixing is very small $\epsilon < 10^{-20}$ \cite{Khan:2024biq, Khan:2025dht}, so that the lifetime of DPDM is long enough compared to the age of the Universe.

The model under consideration has been extensively studied in the literature \cite{Ng:2014iqa, Costa:2022lpy, Khan:2023uii, Khan:2025keb, Khan:2025yoa}. In particular, Ref.~\cite{Khan:2023uii} focuses on constraining the $W_D$ DM from the SM Higgs inflation when it is produced by the freeze-out mechanism. In Ref.~\cite{Khan:2025yoa}, the same object is pursued with the freeze-in mechanism. The vector DM in the context of the low reheating temperature has been explored in Ref.~\cite{Khan:2025keb}. In the current work, we explore the same low-reheating scenario, but we also take into account entropy production, which dilutes the DM abundance. Moreover, we study the production of the vector DM using the freeze-out mechanism \cite{Khan:2023uii} as well in this context. In our DM analysis, we also consider the interactions between the SM and BSM Higgses and $gg$, $\gamma\gamma$, and $Z\gamma$, which are absent at the tree level but generated at the loop level, following Ref.~\cite{Djouadi:2005gi}. Therefore, the present study goes beyond the existing studies and serves as an important addition to the vector DM physics.
The two scalar fields, namely the SM Higgs field and the dark Higgs field, may drive cosmic inflation, when coupled to the Ricci scalar. In Ref.~\cite{Khan:2023uii}, for instance, the SM Higgs inflation scenario has been thoroughly investigated in this model. In the current work, on the contrary, we shall explore the scenario where the dark $U(1)_D$ Higgs field is the inflaton.

\section{Dark matter}
\label{sec:dm}
Let us first discuss DM physics in the model under consideration. While we shall consider the scenario where inflation is driven by the dark Higgs field later in section~\ref{sec:inflation} and study the connection between the dark Higgs inflationary physics and DM physics, our discussion in this section is valid for a generic inflation setup as long as the inflaton potential near the minimum is quadratic. Therefore, for now, we remain agnostic to a concrete realisation of inflation and discuss DM phenomenology considering a generic reheating process.
After the end of inflation, the energy density of the inflaton field gets transferred to the SM thermal bath energy density via processes such as the decay of the inflaton to the SM Higgs, which eventually decays to the SM particles, forming the radiation bath. The exact process depends on a concrete realisation of inflation and the particle physics model that governs the reheating process.

\subsection{Boltzmann equations for DM production in low-reheating scenarios}
In the presence of the inflaton decay with the decay width $\Gamma_{\rm inf}$ during reheating, the energy density of the inflaton $\rho_{\rm inf}$, and the entropy of the Universe $s$, change according to the following differential equations \cite{Gelmini:2006pw}:
\begin{align}
	\frac{d \rho_{\rm inf}}{d t} + \frac{6 k}{k+2} \rho_{\rm inf} &= -\Gamma_{\rm inf} \rho_{\rm inf}
	\,,\\
	\frac{ds}{dt} + 3 s H &= \frac{\Gamma_{\rm inf} \rho_{\rm inf}}{T}
	\,,
\end{align}
where $k$ is the polynomial power in the inflaton potential, $V_{\rm inf}(\varphi) \sim \varphi^k$, $t$ denotes the cosmic time, $T$ represents the temperature, and $H$ is the Hubble parameter. The entropy density is defined as $s(T) = (2 \pi^2/45) g_{s}(T) T^{3}$, where $g_s$ is the effective number of relativistic degrees of freedom for the entropy density. The Hubble parameter is given by $H = \sqrt{(\rho_{\rm inf} + \rho_{R})/(3 M^2_{\rm P})}$, with $M_{\rm P}$ being the reduced Planck mass and $\rho_{R} = (\pi^2 / 30) g_{\rho}(T) T^{4}$ the radiation energy density, where $g_\rho$ is the effective number of relativistic degrees of freedom for the energy density. We assume the initial conditions $\rho_{R}(t_{I}) = 0$ and $\rho_{\rm inf}(t_{I}) = 3 M_{\rm P}^2 H^2_{I}$, where $H_I = H(t=t_I)$ is the Hubble parameter at the end of inflation or, equivalently, at the start of reheating. In addition, we take the quadratic inflaton potential near the minimum, namely $k=2$, in this section. Once we have the radiation bath from the inflaton decay, we can have DM production from the radiation bath through either the freeze-in or freeze-out mechanism.
The Boltzmann equation for the DPDM evolution is given, in terms of the DM number density $n_{W_D}$, by
\begin{align}
	\frac{d n_{W_D}}{d t} + 3 n_{W_D} H = 
	-\langle \sigma v \rangle_{W_{D}W_{D}} 
	\left( n_{W_D}^{2} - n^2_{W_D,{\rm eq}} \right) 
	+ \langle \Gamma \rangle_{A \rightarrow W_D W_D}
	\left(
	n_{A} - n_{A,{\rm eq}} \frac{n_{W_{D}}^2}{n_{W_D,{\rm eq}}^2}
	\right)
	\,,
\end{align}
where $A = h_{1,2}$.
It is more convenient to work with the total particle number $N_{i} \equiv n_{i} a^3$, for $i=\{W_D,A\}$, and the scale factor $a$, in terms of which the above Boltzmann equation becomes
\begin{align}
	\frac{d N_{W_D}}{da} =
	-\frac{\langle \sigma v \rangle_{W_{D}W_{D}}}{H a^{4}} 
	\left( N_{W_D}^{2} - N_{W_D,{\rm eq}}^2 \right) 
	+\frac{ \langle \Gamma \rangle_{A \rightarrow W_{D} W_{D}}}{H a} 
	\left(
	N_{A} - N_{A,{\rm eq}} \frac{N_{W_{D}}^2}{N_{W_D,{\rm eq}}^2}
	\right)
	\,.
    \label{eqn:BE}
\end{align}
Here, we have two terms associated with the DPDM evolution; the first one is associated with the annihilation, and the second one is associated with the decay. When DM is of the WIMP-type, the decay term is insignificant, while for the FIMP-type DM, the decay term would set the DM abundance. In the case of FIMP DM, we also need to multiply the DM density by a factor of two for the pair production of $W_D$ DM. The expressions for the thermal average of cross-section times velocity $\langle \sigma v \rangle_{W_D W_D}$ and the decay width $\langle \Gamma \rangle_{A \rightarrow W_{D}W_{D}}$ can be expressed as follows:
\begin{align}
	\langle \sigma v \rangle_{W_{D}W_{D}} &= \frac{1}{8 M^4_{W_D} T K_{2}^2(M_{W_{D}}/T)} \int_{4 M^2_{W_D}}^{\infty} \sigma_{W_D W_D} \left( s - 4 M^2_{W_D} \right) \sqrt{s} K_{1}\left( \frac{\sqrt{s}}{T} \right)  ds \,,\nonumber\\ 
	\langle \Gamma \rangle_{A \rightarrow W_{D}W_{D}} &= \Gamma_{A \rightarrow W_{D}W_{D}} \frac{K_{1}(M_{W_D}/T)}{K_{2}(M_{W_D}/T)}
	\,,
\end{align}
where $K_{i}(M_{W_D}/T)$ is the modified Bessel function of the second kind of the $i^{\rm th}$ order, and $\sigma_{W_D W_D}$ and $\Gamma_{A \rightarrow W_{D}W_{D}}$ are the DM annihilation cross-section and decay width of particle $A(=h_{1,2})$ to DM $W_D$, which are given in Appendix~\ref{apdx:sigmaGamma}.
Once we obtain the total number of DM particles $N_{W_D}$ after solving Eq.~\eqref{eqn:BE}, we can obtain the yield $Y_{W_D}$ through \cite{Belanger:2024yoj}
\begin{align}
    Y_{W_D} =
    \frac{N_{W_D}}{s(T) a^{3}}
    \,.
\end{align}
With the yield, we can determine the DM relic density using \cite{Edsjo:1997bg}
\begin{align}
    \Omega_{W_D} h^{2} = 2.755\times 10^{8} \times \left( \frac{M_{W_D}}{\rm GeV} \right) \times Y_{W_D}
    \,.
    \label{eqn:DM-density}
\end{align}
It is clear from Eq.~\eqref{eqn:DM-density} that the DM relic density $\Omega_{W_D} h^2$ is proportional to the DM yield $Y_{W_D}$.

It is to be noted that in the present work, we utilise {\tt micrOMEGAs} (v6.2.3)~\cite{Belanger:2001fz} which solves the Boltzmann equation numerically using the freeze-out or freeze-in mechanism, depending on the DM interaction strength with the visible sector and the reheating temperature $T_{\rm reh}$ considered. The characteristic quantity is
\begin{align}
	\Delta \equiv \frac{n_{W_D} - n_{W_D,{\rm eq}}}{n_{W_D,{\rm eq}}}
	\,,
	\label{eqn:WIMP-FIMP-condition}
\end{align}
which indicates how fast the DM number density changes from its equilibrium number density. For a change in the scale factor $\delta a / a < 10^{-3}$, if $|\Delta| < 10^{-2}$, then the DM is called the WIMP-type DM; otherwise, it is called the FIMP-type DM candidate \cite{Belanger:2024yoj}. Thus, when solving the Boltzmann equation, {\tt micrOMEGAs} first checks if DM satisfies the WIMP condition. If the WIMP condition is satisfied, then it tries to find the maximum value of $a$ and starts solving the Boltzmann equation with the DM initial value $z = z_{\rm eq} + \delta z$. If the WIMP condition fails to be satisfied, the Boltzmann equation is solved from $a=1$ with the initial abundance of DM fixed to zero using the freeze-in mechanism.

\begin{figure}[t!]
	\centering
	\includegraphics[angle=0,height=7.5cm,width=7.5cm]{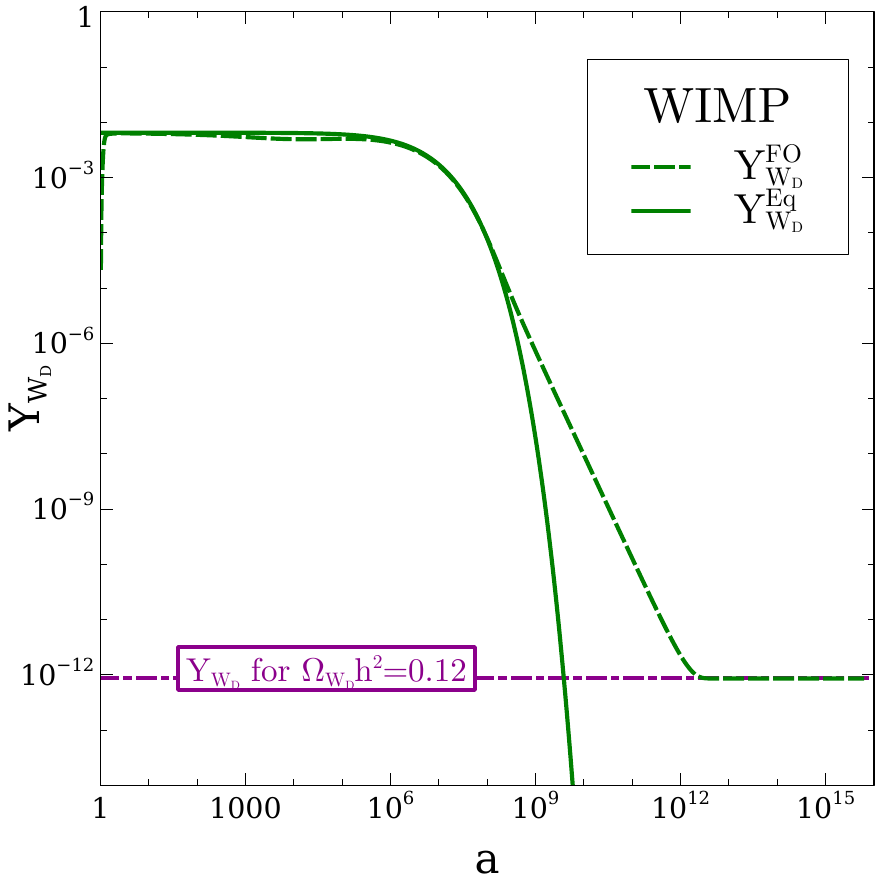}
	\includegraphics[angle=0,height=7.5cm,width=7.5cm]{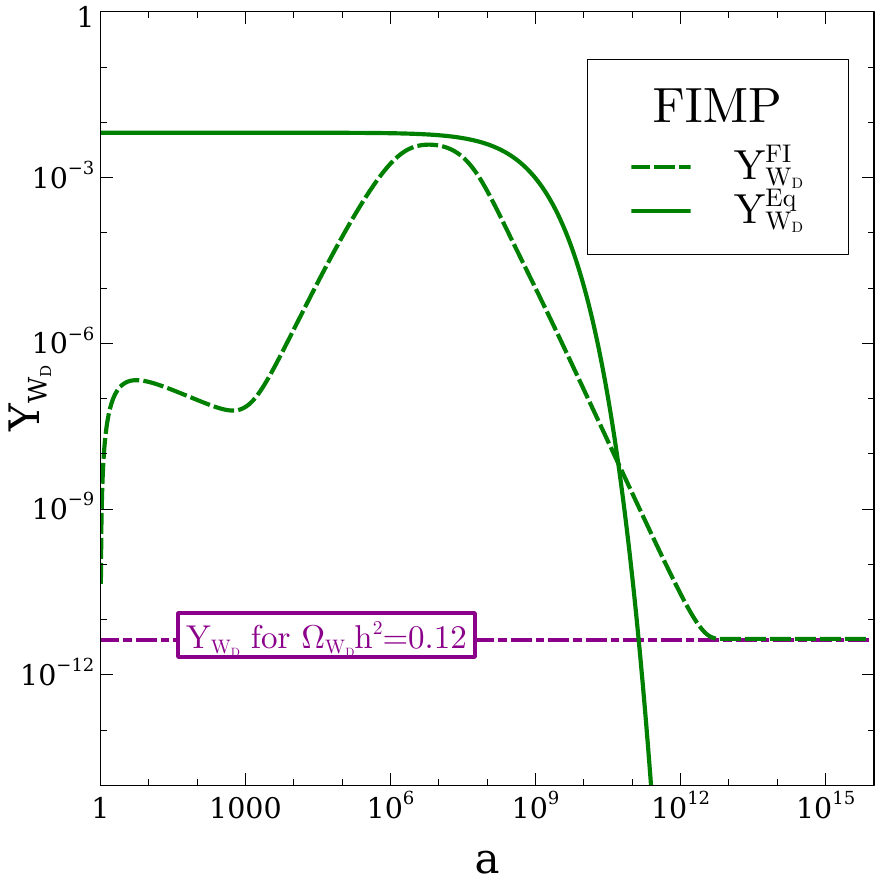}    
	\caption{Evolution of the DM yield $Y_{W_D}$ in terms of the scale factor $a$ for the WIMP-type DM (left) and the FIMP-type DM (right). The green dashed and solid lines represent the DM yield and the equilibrium yield, respectively. The magenta dot-dashed line indicates the yield to produce the correct value of DM relic density from the Planck observation, namely $\Omega_{W_D} h^2 = 0.12$. The parameters are chosen as follows: $g_{D} = 0.2$, $M_{W_D} = 500$ GeV, $M_{h_2} = 300$ GeV, $\sin\alpha = 0.1$, and $\Gamma_{\rm inf} = 1.05 \times 10^{-17}$ GeV for the left panel, and $g_{D} = 10^{-5}$, $M_{W_D} = 100$ GeV, $M_{h_2} = 900$ GeV, $\sin\alpha = 0.1$, and $\Gamma_{\rm inf} = 4.4 \times 10^{-18}$ GeV for the right panel.}
	\label{fig:line-plot-0}
\end{figure}

In Fig.~\ref{fig:line-plot-0}, we show the variation of the DM yield $Y_{W_D}$ with the scale factor $a$ for the WIMP-type DM (left) and the FIMP-type DM (right). The green dashed and solid lines represent the DM yield and the equilibrium yield, respectively, while the magenta dot-dashed line marks the yield for the correct value of DM relic density from the Planck observation, namely $\Omega_{W_D} h^2 = 0.12$. From the left panel of Fig.~\ref{fig:line-plot-0}, one may see that, depending on the condition of $\Delta < 10^{-2}$, DM is initially in thermal equilibrium and closely follows the equilibrium yield as depicted by the dashed and solid lines at early times. As the Universe evolves, DM goes out of equilibrium, {\it i.e.}, the DM evolution deviates from the equilibrium evolution, and continues its density reduction due to entropy production and freezes out to a particular value when the inflaton fully decays and no more entropy change happens. On the other hand, as presented in the right panel of Fig.~\ref{fig:line-plot-0}, if the DM production is through the freeze-in mechanism, DM never reaches thermal equilibrium. In this case of the FIMP-type DM, DM is maximally produced from the thermal bath and then starts reducing its yield due to entropy production from the decay of the inflaton. When the inflaton decays fully, DM yield freezes into a particular value depending on the model parameters. In both left and right panels of Fig.~\ref{fig:line-plot-0}, we clearly see how DM evolves compared to its equilibrium density with the scale factor depending on the DM type. The evolution of the DM relic density $\Omega_{W_D} h^2$ then follows the same evolution behaviour as noted by Eq.~\eqref{eqn:DM-density}.

\begin{figure}[t!]
	\centering
	\includegraphics[angle=0,height=6.5cm,width=7.5cm]{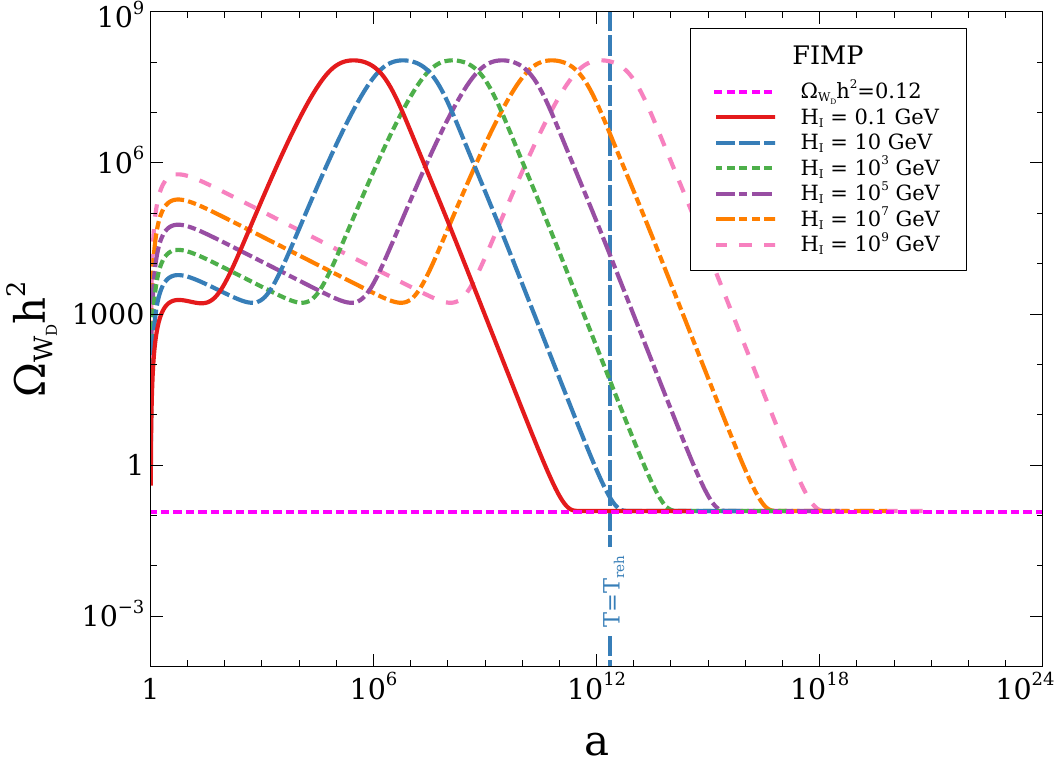}
	\includegraphics[angle=0,height=6.5cm,width=7.5cm]{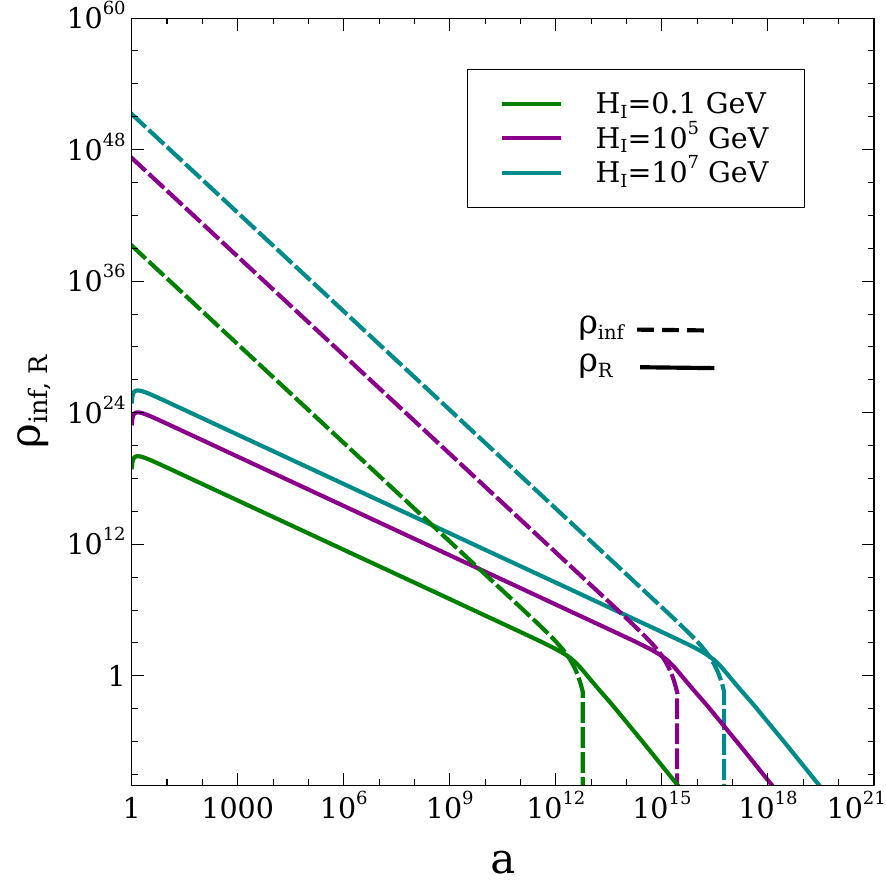}    
	\caption{Evolution of the DM relic density $\Omega_{W_D}h^2$ (left) and the inflaton energy density $\rho_{\rm inf}$ and radiation energy density $\rho_R$ (right) in terms of the scale factor $a$ for different values of Hubble parameter $H_I$ at the start of reheating. The other parameters are chosen as $g_{D} = 10^{-5}$, $M_{W_D} = 100$ GeV, $M_{h_2} = 900$ GeV, $\sin\alpha = 0.1$, and $\Gamma_{\rm inf} = 4.4 \times 10^{-18}$ GeV. DM is produced by the freeze-in mechanism obeying the condition $|\Delta| > 10^{-2}$; see the text below Eq.~\eqref{eqn:WIMP-FIMP-condition}.}
	\label{fig:line-plot-1}
\end{figure}

\subsection{Evolution of the DM relic density for WIMP and FIMP cases}
The left panel of Fig.~\ref{fig:line-plot-1} presents the variation in the DM relic density for different values of the Hubble parameter $H_I$ at the end of inflation. The vertical dashed line indicates the scale factor at which the temperature becomes the reheating temperature for $H_I = 10$ GeV, while the horizontal dashed line represents $\Omega_{W_D}h^2 = 0.12$. We can see that for different values of $H_I$, the DM relic density, which is produced by the freeze-in mechanism, hits its maximum value at different values of the scale factor $a$ and then decreases due to the dilution from the continuous production of entropy until the reheating temperature $T = T_{\rm reh}$. The final DM relic is, however, independent of the initial condition $H_I$, indicating the infrared production of DM. The thermal bath is formed from the inflaton decay, and then the DM production occurs. We note that the DM production directly from the inflaton decay is assumed to be absent which can be achieved by demanding $g_{D} > \sqrt{\lambda_{D}}$; we shall discuss this in more detail shortly. We have seen that for higher values of $H_I$, {\tt micrOMEGAs} takes a longer time to execute. Since the final amount of DM is independent of the choice of $H_I$, in our numerical analysis, we consider a low value of $H_I$ which reduces the runtime of {\tt micrOMEGAs} significantly, not affecting the final DM relic density. We have also checked the convergence of the DM relic curves when plotted with $z = M_{W_D}/T$ for the $x$-axis. The right panel of Fig.~\ref{fig:line-plot-1} shows the variation of the inflaton energy density $\rho_{\rm inf}$ (solid lines) and the radiation energy density $\rho_{R}$ (dashed lines) with the scale factor $a$ for three different values of the Hubble parameter $H_I$ at the start of reheating. We observe that, depending on the initial condition of the inflaton energy density, which is set by the initial value of the Hubble parameter, the onset of the radiation-dominated era occurs at different values of the scale factor. The slope change in the lines corresponds to the transition from the inflaton-dominated era to the radiation-dominated era. We can clearly see that the transition occurs later for a lower initial Hubble parameter.

\begin{figure}[t!]
	\centering
	\includegraphics[angle=0,height=7cm,width=7.5cm]{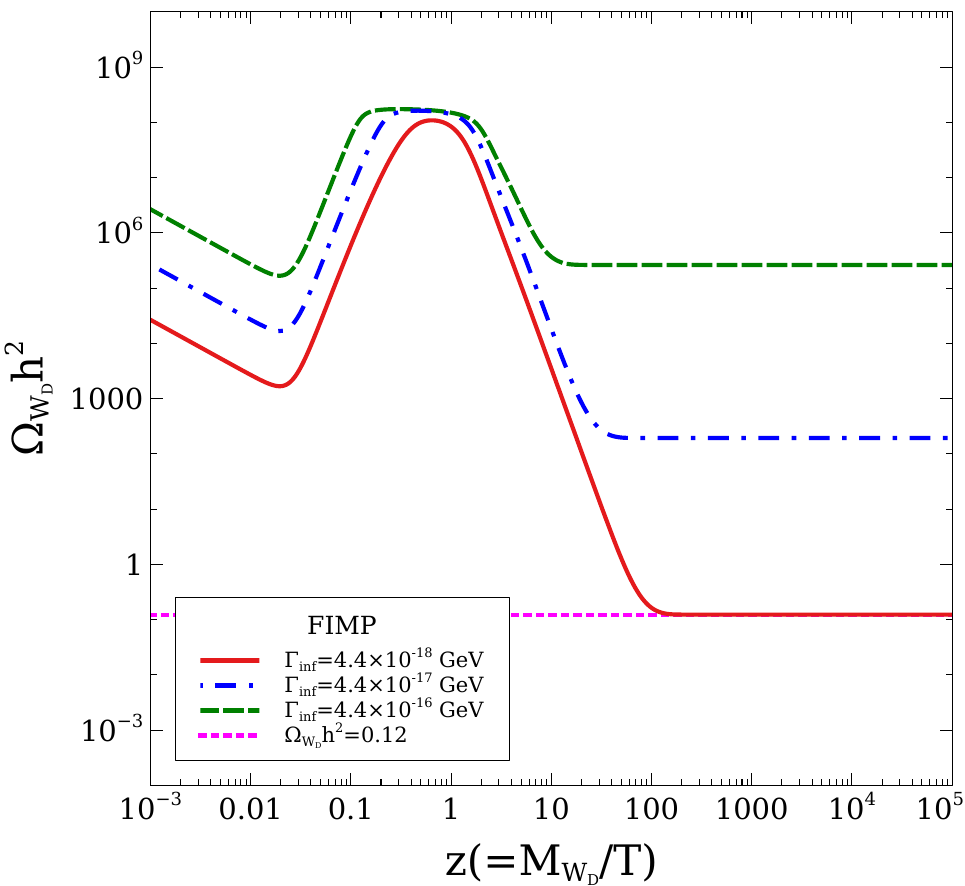}
	\includegraphics[angle=0,height=7cm,width=7.5cm]{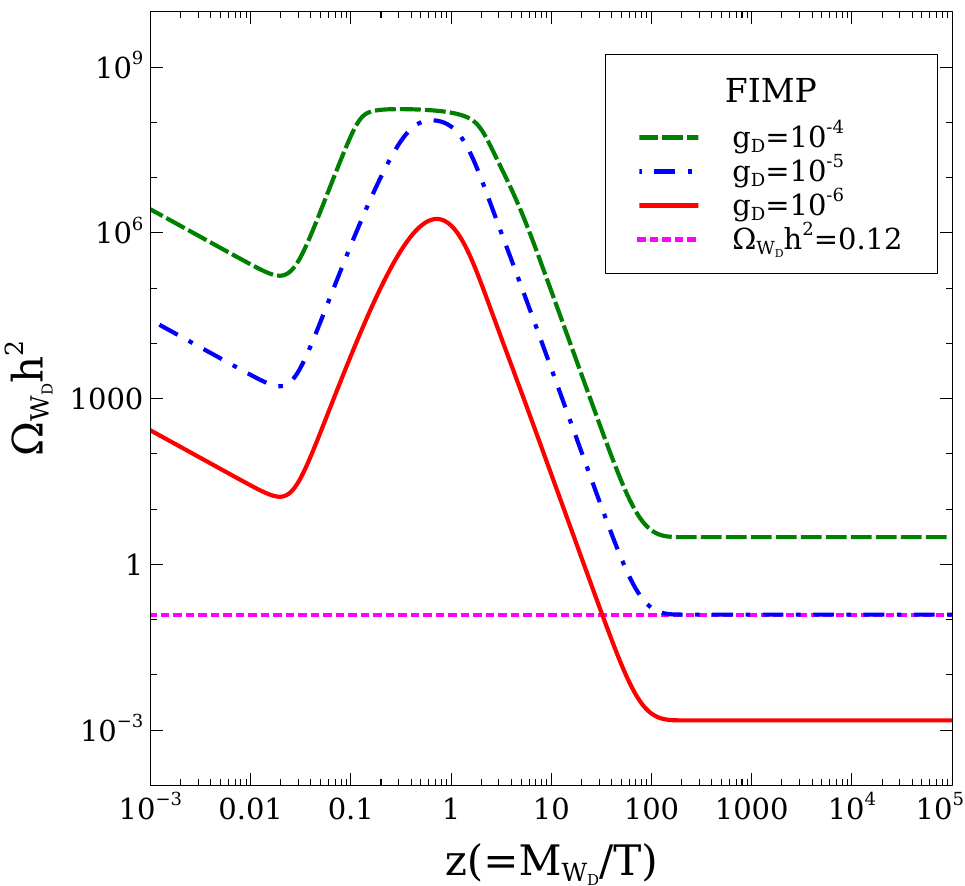}
	\caption{Evolution of the DM relic density $\Omega_{W_D}h^2$ in term of $z=M_{W_D}/T$ for different values of the inflaton decay width $\Gamma_{\rm inf}$ (left) and the dark gauge coupling $g_D$ (right). The horizontal magenta line indicates $\Omega_{W_D}h^2 = 0.12$. DM is produced by the freeze-in mechanism, and the rest of the parameters are chosen to be the same as those shown in Fig.~\ref{fig:line-plot-1}.} 
	\label{fig:line-plot-2}
\end{figure}

In the left panel of Fig.~\ref{fig:line-plot-2}, we have the evolution of the DM relic density $\Omega_{W_D}h^2$ in term of $z=M_{W_D}/T$ for three different values of the inflaton decay width $\Gamma_{\rm inf}$. The right panel of Fig.~\ref{fig:line-plot-2}, on the other hand, shows the evolution of the DM relic density $\Omega_{W_D}h^2$ for three different values of the dark gauge coupling $g_D$.
For both cases, DM is produced by the freeze-in mechanism, obeying the condition $|\Delta| > 10^{-2}$. One of the dominating channels for the DM production is the decay of the dark Higgs $h_2$ to DM $W_D$. As the inflaton decay width $\Gamma_{\rm inf}$ increases, we have higher values of the reheating temperature $T_{\rm reh}$, and hence, less suppression from the $e^{-M_{W_{D}, h_{2}}/T_{\rm reh}}$ factor, which produces more DM as seen from the figure. Moreover, as $T_{\rm reh}$ increases, there is less suppression from the entropy production, and thus, we see that DM freezes in with a higher abundance. We can see a plateau around $z = 0.1 - 1$, which implies that the DM abundance has already reached the equilibrium. From the right panel of Fig.~\ref{fig:line-plot-2}, we see that the DM production increases as the dark gauge coupling $g_D$ increases. This is consistent with the freeze-in mechanism. We observe that the DM relic abundance gets diluted until the same point, $z\simeq 100$, as the inflaton decay width, or, equivalently, the reheating temperature, is fixed, but it freezes in to a different value depending on the amount of production. From Fig.~\ref{fig:line-plot-2}, we find that varying $\Gamma_{\rm inf}$ and $g_D$ may put the DM abundance in a suitable range. In particular, if we change $\Gamma_{\rm inf}$, then it does not affect the DM interaction, and hence, we expect no effect on the detection prospect of the DM. Based on this observation, we will not multiply DM direct/indirect detection by its fraction because we can find a suitable value of $\Gamma_{\rm inf}$ which provides the 100\% DM without affecting the direct/indirect detection cross section.

\begin{figure}[t!]
	\centering
	\includegraphics[angle=0,height=7cm,width=7.5cm]{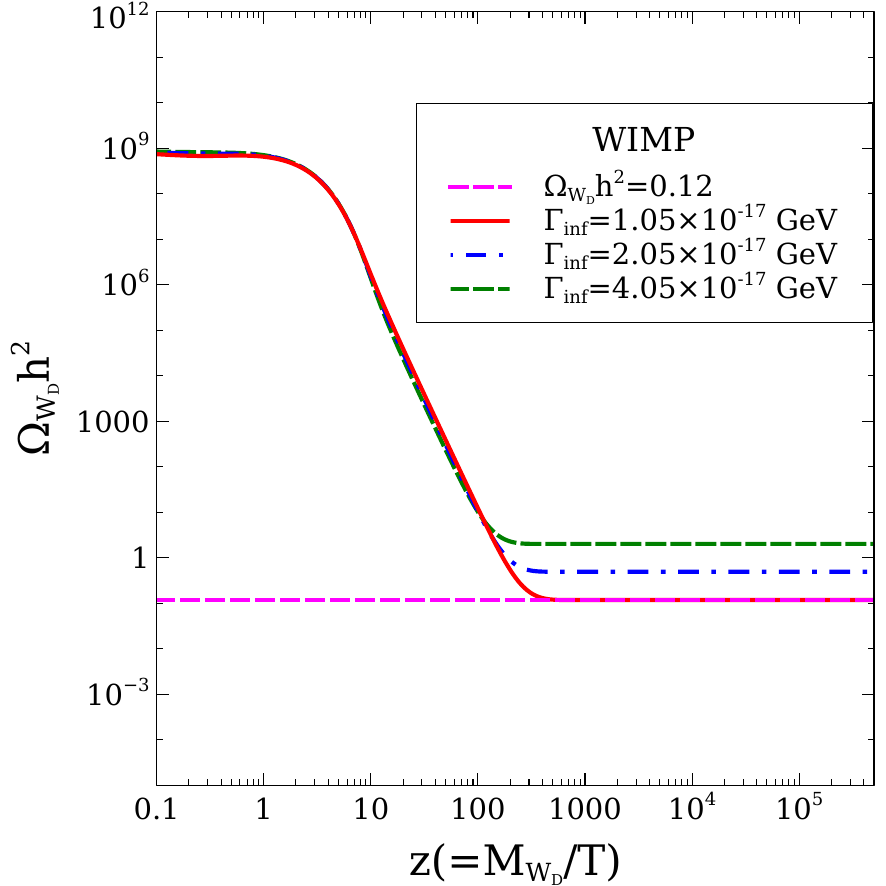}
	\includegraphics[angle=0,height=7cm,width=7.5cm]{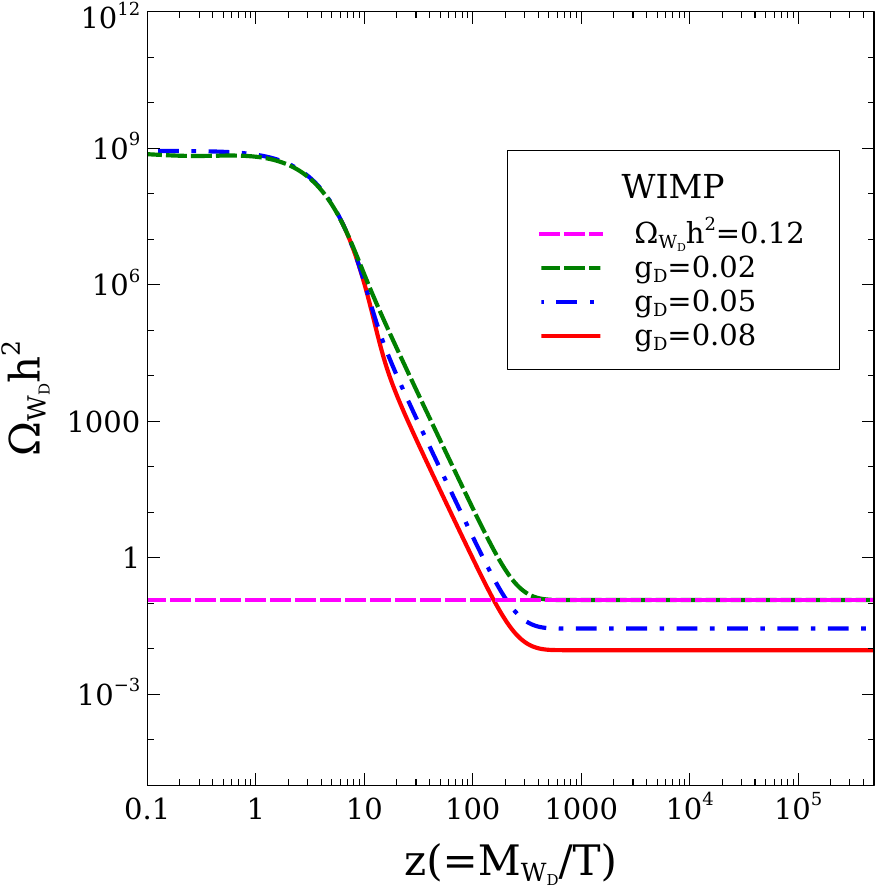}
	\caption{Evolution of the DM relic density $\Omega_{W_D}h^2$ in term of $z=M_{W_D}/T$ for different values of the inflaton decay width $\Gamma_{\rm inf}$ (left) and the dark gauge coupling $g_D$ (right) just like in Fig.~\ref{fig:line-plot-2}. Unlike Fig.~\ref{fig:line-plot-2}, however, the DM is produced by the freeze-out mechanism this time, so we have WIMP-type DM here. The other parameters are chosen as $\sin\alpha = 0.1$, $M_{h_2} = 300$ GeV, $M_{W_D} = 500$ GeV, $g_{D} = 0.02$ and $\Gamma_{\rm inf} = 1.05\times 10^{-17}$ GeV.  The horizontal magenta line indicates $\Omega_{W_D}h^2 = 0.12$.} 
	\label{fig:line-plot-3}
\end{figure}

In Fig.~\ref{fig:line-plot-3}, the evolution of the DM relic density is shown for the WIMP-type DM. In the left panel, three different values of the inflaton decay width $\Gamma_{\rm inf}$ are considered, whereas in the right panel, we have three different values of the dark gauge coupling $g_D$. From the left panel, we see that higher values of $\Gamma_{\rm inf}$, which correspond to higher reheating temperatures, lead to more abundant DM. This is mainly because for lower values of $T_{\rm reh}$, DM abundance is diluted for a longer time due to the entropy production. From the right panel, we observe the standard WIMP-type DM behaviour, {\it i.e.}, $\Omega_{W_{D}}h^{2} \propto 1/\langle \sigma v \rangle$. As we increase the value of $g_D$, we have a lower DM abundance as $\langle \sigma v \rangle$ becomes larger. All three values of $g_D$ experience the same amount of entropy dilution due to the fixed value of the reheating temperature or, equivalently, the inflaton decay width.

\begin{figure}[t!]
	\centering
	\includegraphics[angle=0,height=7cm,width=7.5cm]{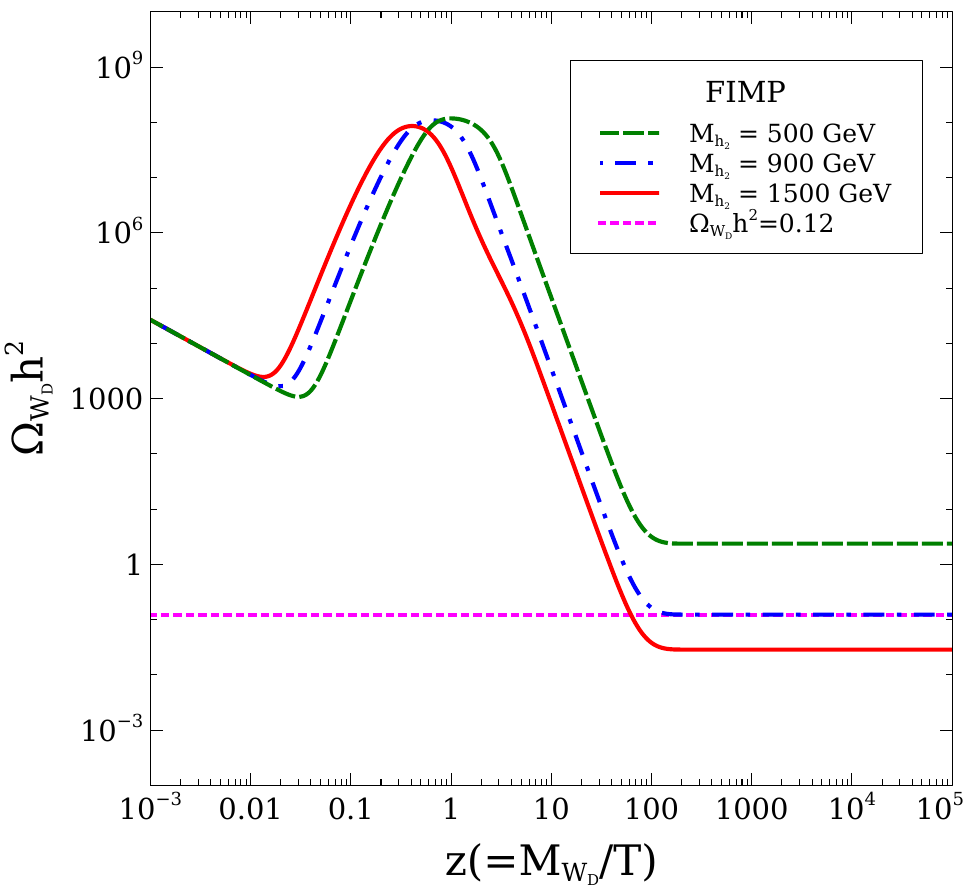}
	\includegraphics[angle=0,height=7cm,width=7.5cm]{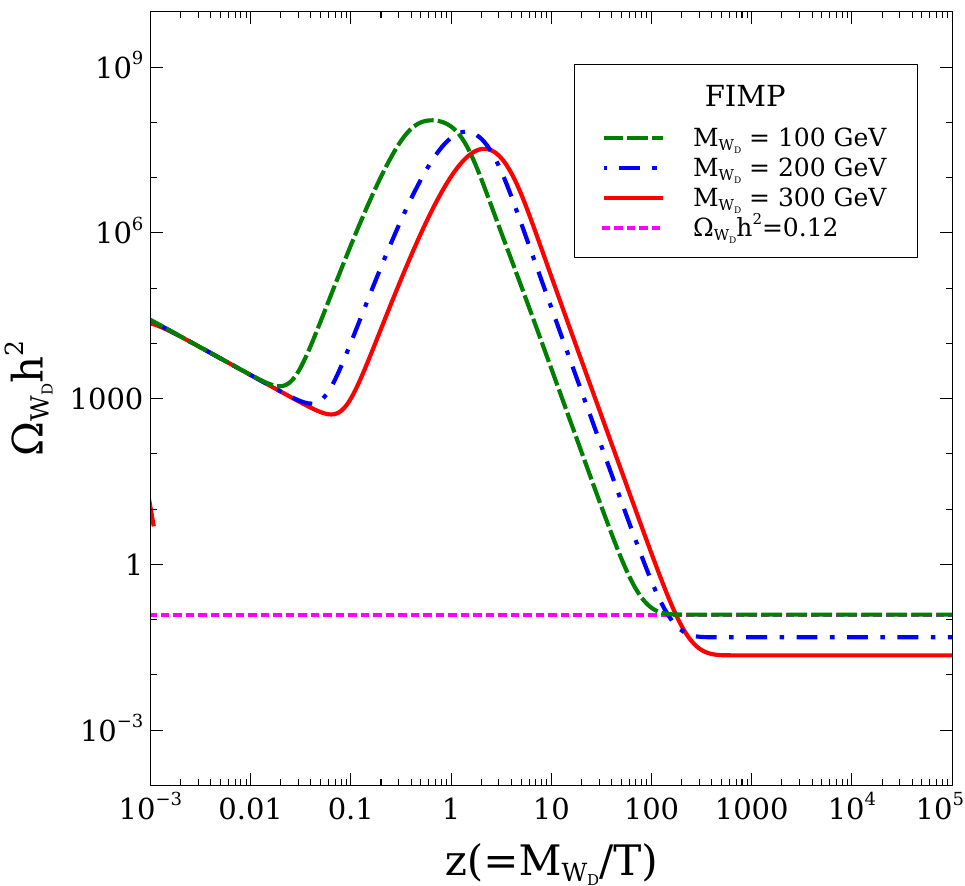}
	\caption{Evolution of the DM relic density $\Omega_{W_D}h^2$ in term of $z=M_{W_D}/T$ for different values of the dark Higgs mass $M_{h_2}$ (left) and the DPDM mass $M_{W_D}$ (right). The horizontal magenta line indicates $\Omega_{W_D}h^2 = 0.12$. The rest of the parameters are chosen to be the same as those shown in Fig.~\ref{fig:line-plot-1}.}
	\label{fig:line-plot-4}
\end{figure}

The dependence of the DM relic density on the dark Higgs mass $M_{h_2}$ and the DPDM mass $M_{W_D}$ is presented in Fig.~\ref{fig:line-plot-4}. From the left panel, we see the production of DM up to $z \simeq 1$, followed by the dilution until $z \simeq 100$. After $z \simeq 100$, the DM density gets frozen. Before the freeze-in, DM is produced during the inflaton-dominated epoch where the temperature $T(a) \propto a^{-3/8}$ instead of $T(a) \propto a^{-1}$. Therefore, the production of DM does not vary as in the standard radiation case. We can see that the DM relic density reaches roughly the same maximum value for all the three values of the dark Higgs mass before the dilution starts. The DM production happens more actively up to $T < M_{h_2}$, after which the production gets Boltzmann-suppressed. This behaviour is directly shown in the figure; for higher values of the dark Higgs mass, the DM production peaks at lower values of $z$. For all the three values, the reheating temperature is fixed which makes the DM abundance diluted for longer. As a result, higher values of the dark Higgs mass correspond to lower values of the DM abundance. From the right panel, we observe that the DM relic density varies as $\Omega_{W_D} h^{2} \propto g^2_{D}/M_{W_D}$.

\begin{figure}[t!]
	\centering
	\includegraphics[angle=0,height=7cm,width=7.5cm]{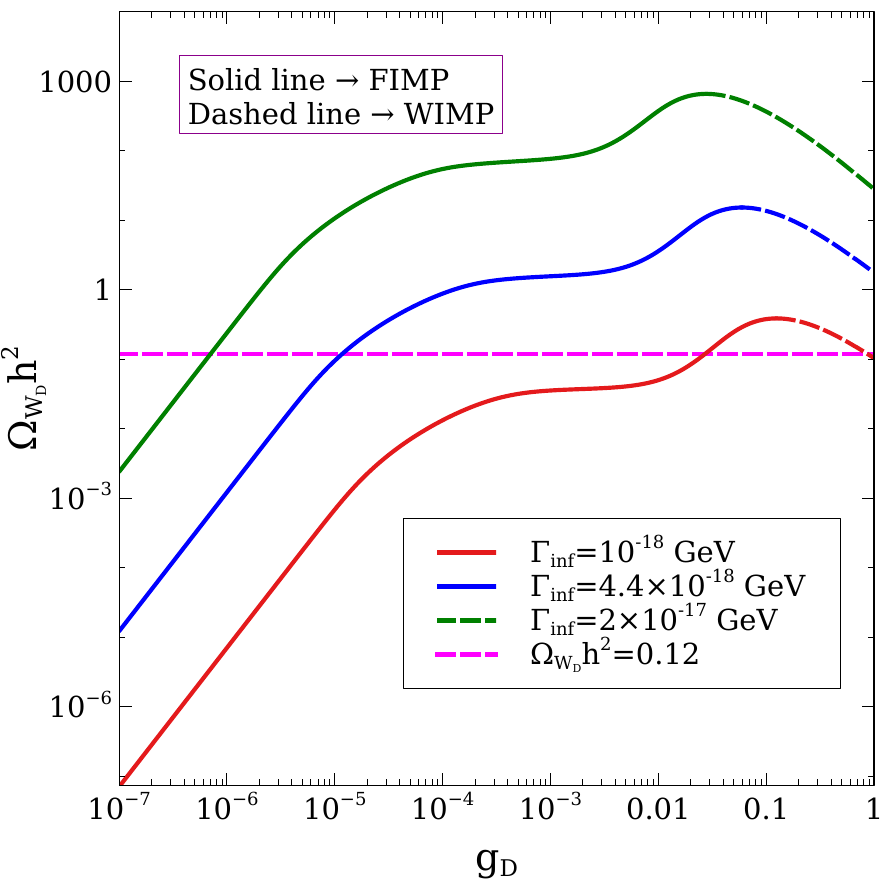}
	\includegraphics[angle=0,height=7cm,width=7.5cm]{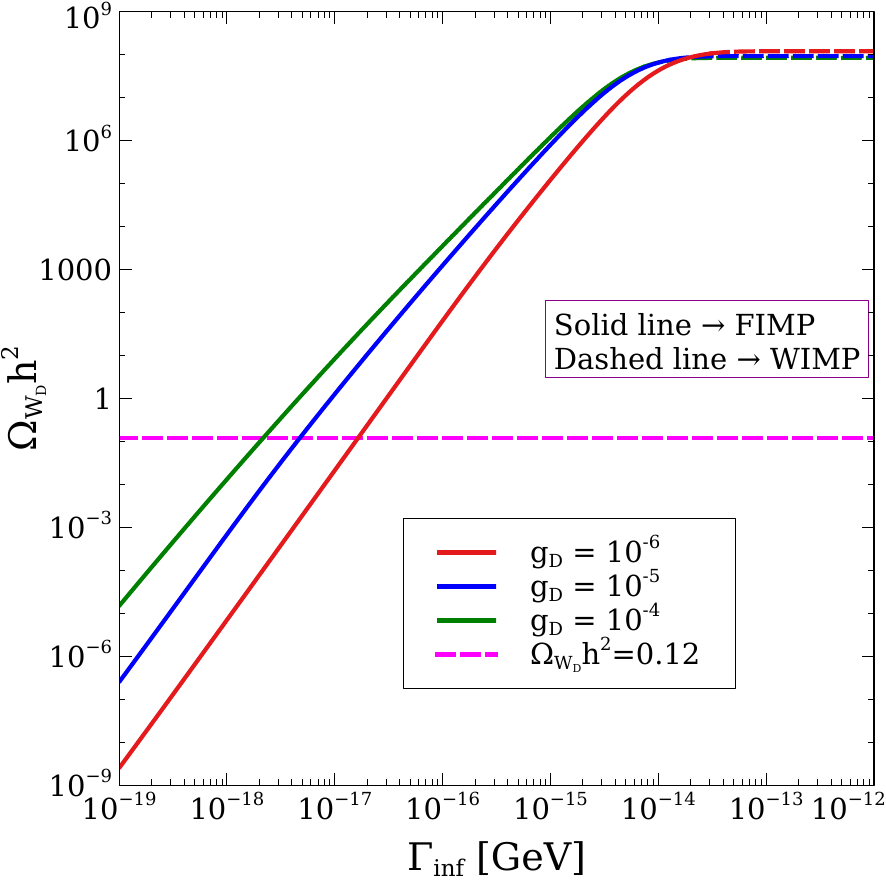}
	\caption{Left panel: Evolution of the DM relic density $\Omega_{W_D}h^2$ in terms of the dark gauge coupling $g_D$ for three different values of the inflaton decay width $\Gamma_{\rm inf}$. Right panel: Evolution of the DM relic density $\Omega_{W_D}h^2$ in terms of the inflaton decay width $\Gamma_{\rm inf}$ for three different values of the dark gauge coupling $g_D$. The horizontal magenta line indicates $\Omega_{W_D}h^2 = 0.12$. The solid and dashed lines represent the DM production by the freeze-in and freeze-out mechanisms, respectively. The rest of the parameters are chosen to be the same as those shown in Fig.~\ref{fig:line-plot-1}.}
	\label{fig:line-plot-5}
\end{figure}

In the left panel of Fig.~\ref{fig:line-plot-5}, we present the DM relic density variation with respect to the dark gauge coupling $g_D$ for three fixed values of the inflaton decay width $\Gamma_{\rm inf}$, while in the right panel, we have the DM relic density variation with respect to the inflaton decay width $\Gamma_{\rm inf}$ for three fixed values of the dark gauge coupling $g_D$. The solid lines mean DM produced by the freeze-in mechanism, while the dashed lines indicate DM produced by the freeze-out mechanism. From the left panel, we can see that the DM production happens following the relation $\Omega_{W_{D}}h^{2} \propto g^2_{D}/M_{W_D}$. As we increase the $g_D$ value, we first have the growth of the DM production, and after a certain value of the $g_D$, DM satisfies the equilibrium condition depicted by the dashed line. With the increment of $\Gamma_{\rm inf}$, we observe that DM achieves the equilibrium condition for a lower value of $g_D$; this can be understood by noting that DM gets diluted less for higher values of $\Gamma_{\rm inf}$, and thus, it is easier to satisfy the equilibrium condition earlier. Once DM reaches thermal equilibrium, we see that the DM relic density falls as $g_D$ increases, which is precisely the case for the WIMP-type DM. From the right panel, we see the linear production of DM as $\Gamma_{\rm inf}$ increases. As we have noted earlier, this happens because DM is less diluted for higher values of $\Gamma_{\rm inf}$ or, equivalently, higher values of the reheating temperature. After a certain value of $\Gamma_{\rm inf}$, DM enters thermal equilibrium and freezes out relativistically because of the choice of the small dark gauge coupling. Therefore, we do not see much variation in the DM relic density for different values of $g_D$.

\begin{figure}[t!]
	\centering
	\includegraphics[angle=0,height=7cm,width=7.5cm]{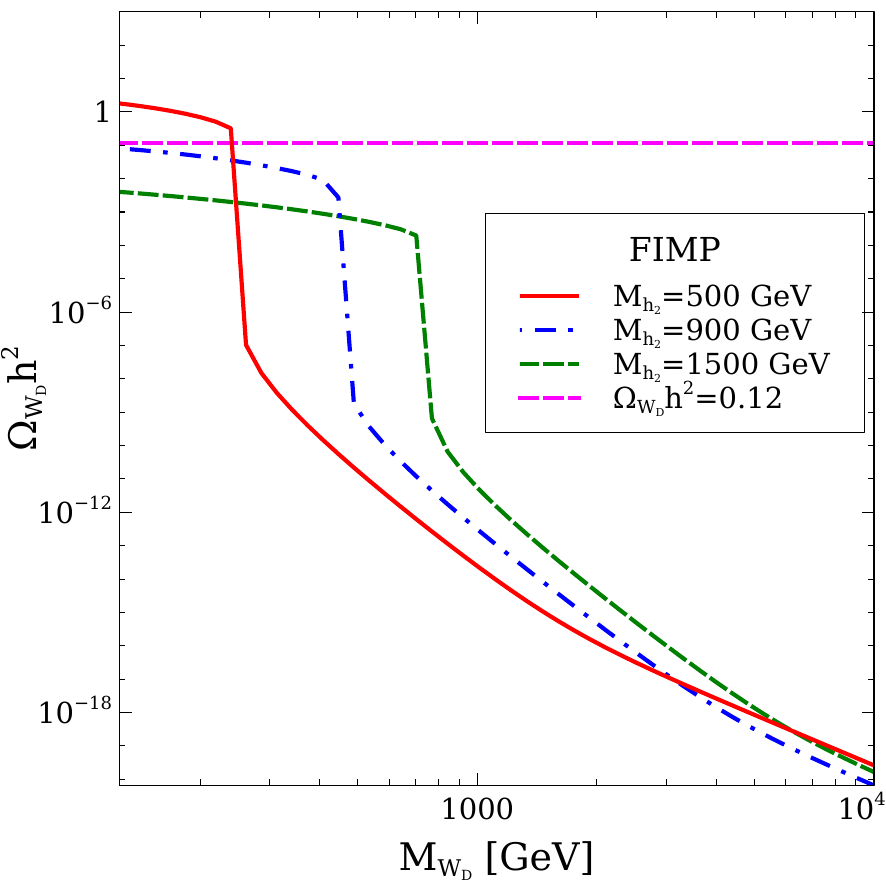}
	\includegraphics[angle=0,height=7cm,width=7.5cm]{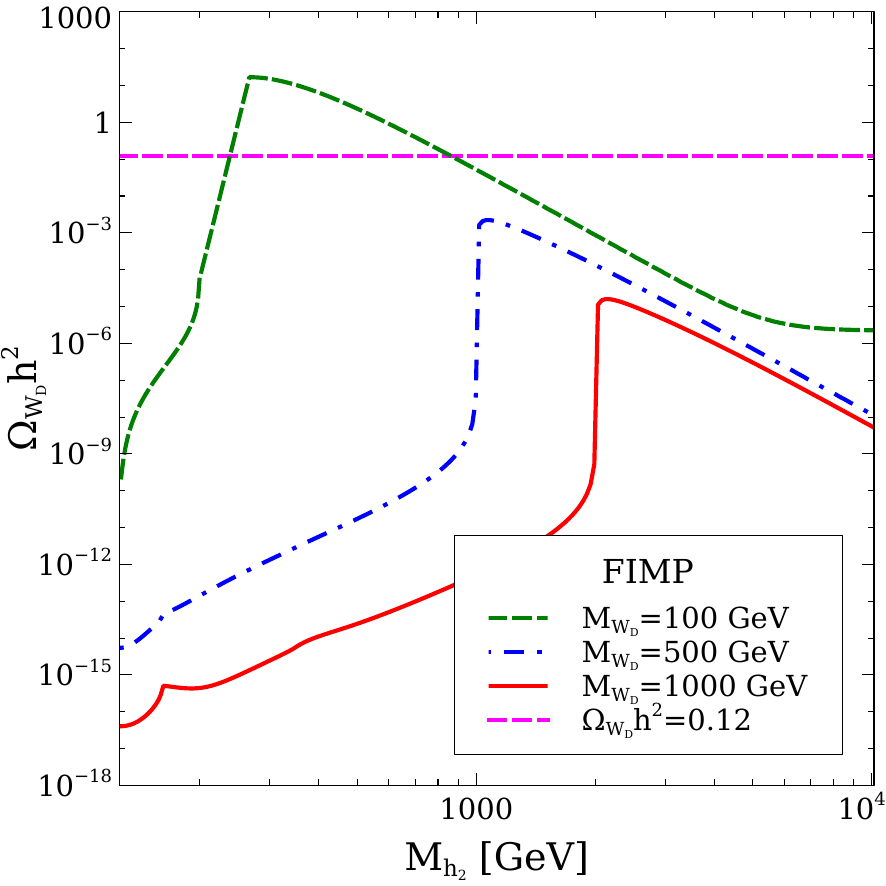}
	\caption{Left panel: Evolution of the DM relic density $\Omega_{W_D}h^2$ in terms of the DPDM mass $M_{W_D}$ for three different values of the dark Higgs mass $M_{h_2}$. Right panel: Evolution of the DM relic density $\Omega_{W_D}h^2$ in terms of the dark Higgs mass $M_{h_2}$ for three different values of the DPDM mass $M_{W_D}$. The horizontal magenta line indicates $\Omega_{W_D}h^2 = 0.12$. The rest of the parameters are chosen to be the same as those shown in Fig.~\ref{fig:line-plot-1}.}
	\label{fig:line-plot-6}
\end{figure}

The left panel of Fig.~\ref{fig:line-plot-6} shows the variation of the DM relic density with respect to the DPDM mass $M_{W_D}$ for three fixed values of the dark Higgs mass $M_{h_2}$. Until $M_{W_D} < M_{h_2}/2$, we have the decay contribution, and the DM relic density slowly decreases. At the threshold region, we have a sudden drop, and then the DM production is mainly dominated by the annihilation process $h_{2} h_{2} \rightarrow W_{D} W_{D}$. In the annihilation-dominated region, the DM production mainly varies as $\Omega_{W_{D}} h^{2} \propto 1/M^2_{W_{D}}$ which is also clearly seen from the figure. As the dark Higgs mass is larger, DM peaks at a higher temperature, having more time to get diluted due to the entropy production, leading to a lower abundance when the DM is produced by the decay. However, when it is produced by the annihilation and for $M_{W_D} > M_{h_2}$, the production is mainly controlled by $M_{W_D}$ and independent of the dark Higgs mass. This is because for higher values of the DPDM mass, the thermally-averaged cross-section is monitored by the DPDM mass. In the right panel, we have the variation of the DM relic density with respect to the dark Higgs mass. When the process is dominated by the annihilation, we see that the DM relic increases linearly with $M_{h_2}$ because the DM relic density varies with $M_{h_2}$ in a power-law. After that, there is a sharp rise in the production when the $h_{2}\rightarrow W_D W_D$ production modes open. We then see a linear fall in the DM relic density as the $M_{h_2}$ value increases. This is again due to more dilution of the DM density from the entropy production during reheating.

\subsection{Relevant bounds for the DM study}
Having observed the tendency of the DM relic density with respect to various model parameters, we are now in a position to perform a thorough parameter scan. Let us first discuss relevant constraints:
\begin{itemize}
	\item Collider bounds
	
	In the present study, the relevant collider bounds come from the measurement of the Higgs signal strength and its branching ratio. The SM Higgs signal strength gives us a bound on the Higgs mixing angle $\sin\alpha$. The Higgs invisible decay data can easily be evaded by making the DPDM mass larger than the half of the SM Higgs mass. A lower DPDM mass region could also be considered by adjusting the dark gauge coupling and/or the Higgs mixing angle. The bound on the Higgs mixing angle is given by \cite{ATLAS:2016neq, Herner:2016woc, ATLAS:2022vkf, CMS:2022dwd, Heo:2024cif, Khan:2025yko, Covi:2025erx}
	\begin{align}
		\sin\alpha \leq 0.23 ~{\rm at}~ 2\sigma\,.
	\end{align}
	The bound on the SM Higgs branching ratio to invisible modes can be taken as \cite{ATLAS:2022yvh}
	\begin{align}
		{\rm Br}_{h_{1} \rightarrow {\rm inv}} \leq 0.103 ~{\rm at}~ 95\% ~{\rm C.L.}\,.
	\end{align}
	We use {\tt micrOMEGAs} to determine the Higgs invisible branching ratio. As we will see in section~\ref{subsec:reheating}, in our study we require smaller values of the mixing quartic coupling to realise the low-reheating scenario, which naturally evades the bounds associated with the Higgs data.
	
	\item DM direct detection
	
	The cross-section for the DM direct detection process $W_{D} N \rightarrow W_{D} N$, with $N$ being a nucleon, in the model under consideration is given by
	\begin{align}
		\sigma_{\rm SI} = \frac{\mu^2 \sin^2(2\theta) g_D^2}{4 \pi v_H^2}
		\left( \frac{1}{M_{h_1}^2} - \frac{1}{M_{h_2}^2} \right)^2
		\left[ \frac{Z f^p_\alpha  + (A-Z) f^n_\alpha}{A} \right]^2
		\,,\label{eqn:dd-cs}
	\end{align}
	where $\mu = M_{W_D} M_N / (M_{W_D} + M_N)$ is the reduced mass consisting of the nucleon mass $M_N$ and the DPDM mass $M_{W_D}$, and $Z$ ($A$) is the atomic number (total mass of atom). The form factors associated with the proton and neutron can be expressed as
	\begin{align}
		f^p_\alpha &= M_N \left[
		\frac{7}{9} \left(f_p^u + f_p^d + f_p^s\right) + \frac{2}{9}
		\right]
		\,,\nonumber\\
		f^n_\alpha &= M_N \left[
		\frac{7}{9} \left(f_n^u + f_n^d + f_n^s\right) + \frac{2}{9}
		\right]
		\,,
	\end{align}
	with $f_p^u = f_n^d = 0.02$, $f_n^u = f_p^d = 0.026$, and $f_p^s = f_n^s = 0.043$ \cite{Junnarkar:2013ac}. 
	
	\item DM indirect detection
	
	The DM indirect detection in the present work is governed by the process $W_{D} W_{D} \rightarrow W^{+} W^{-}$ whose cross-section in the nonrelativistic limit is given by
	\begin{align}
		\left\langle \sigma v \right\rangle_{WW} &\simeq 
		\frac{4 M^4_{W_{D}} - 4 M^2_{W_{D}} M^2_{W} + 3 M^4_{W} }{96\pi M^2_{W_{D}} M^4_{W}}
		\sqrt{1 - \frac{M_{W}^2}{M_{W_{D}}^2}}
		\nonumber\\&\qquad\times
		\left\vert
		\sum_{i=1,2} \frac{g_{h_{i} W_{D} W_{D}} g_{h_{i} W W}}{(4 M_{W_{D}}^2 - M_{h_{i}}^2) + i \Gamma_{h_{i}} M_{h_{i}}} \right\vert^2
		\,.\label{eqn:ID-cs}
	\end{align}
	Here, the vertex factors $g_{h_{i} W_{D} W_{D}}$ and $g_{h_{i} W W}$ ($i=1,2$) take
	\begin{align}
		g_{h_{1(2) W_{D}W_{D}}} &= - 2 g_{D} M_{W_{D}} \sin\theta (-\cos\theta)
		\,,\\
		g_{h_{1(2) WW}} &= \frac{e^{2} v_H }{2 \sin^{2}{\theta_w}} \cos\theta (\sin\theta)
		\,,
	\end{align}
	For the determination of the decay widths, $\Gamma_{h_1}$ and $\Gamma_{h_2}$, {\tt micrOMEGAs} is utilised.
\end{itemize}
On top of these, when obtaining the allowed points, we demand that the DM relic density falls in the range,
\begin{align}
	10^{-5} \leq \Omega_{W_{D}} h^{2} \leq 0.1284\,.
\end{align}

Before we present results of the parameter scan, let us explore the possibility of realising inflation in the same setup.

\section{Dark Higgs inflation}
\label{sec:inflation}

\subsection{General discussions on dark Higgs inflation}
So far, we have been agnostic to a concrete realisation of inflation and discussed DM phenomenology, focusing on the quadratic inflaton potential near the minimum during reheating and treating the reheating temperature $T_{\rm reh}$ or, equivalently, the inflaton decay width $\Gamma_{\rm inf}$ as a free parameter.
However, as the DPDM model under consideration contains two scalar fields, namely the SM Higgs field and the dark Higgs field, which could drive inflation, we may realise inflation as well in the same model. In Ref.~\cite{Khan:2023uii}, for example, an inflationary scenario where the SM Higgs field plays the role of the inflaton has been investigated in detail. In this work, we explore the possibility of having the dark $U(1)_D$ Higgs field as the inflaton field.

It is well known that inflation with a power-law scalar potential is incompatible with CMB observations \cite{Planck:2018jri} such as the scalar spectral index $n_s$ and the tensor-to-scalar ratio $r$. However, if one allows a coupling between the inflaton field and gravity through the so-called nonminimal coupling term of the form $\varphi^2 R$, the inflationary model becomes compatible with observations \cite{Futamase:1987ua, Fakir:1990eg, Cervantes-Cota:1995ehs, Komatsu:1999mt, Bezrukov:2007ep, Park:2008hz}. We note that the presence of such a nonminimal coupling term is seen natural as it has the mass dimension of four, and there is no symmetry argument that forbids it. Including the nonminimal couplings between the two Higgs fields and gravity, and focusing on the scalar sector which is sufficient for the study of inflation, the action is given by
\begin{align}
	S &= \int d^4x \, \sqrt{-g_{\rm J}} \, \Bigg[
	\frac{M_{\rm P}^2}{2}\left(
	1 + \xi_H \frac{h^2}{M_{\rm P}^2} + \xi_D \frac{\phi^2}{M_{\rm P}^2}
	\right)R_{\rm J}
	\nonumber\\&\qquad\qquad\qquad\qquad
	-\frac{1}{2}g_{\rm J}^{\mu\nu}\partial_\mu h \partial_\nu h
	-\frac{1}{2}g_{\rm J}^{\mu\nu}\partial_\mu \phi \partial_\nu \phi
	-V_{\rm J}(h,\phi)
	\Bigg]\,,
\end{align}
where the subscript J refers to the Jordan frame, $h$ and $\phi$ denote the SM Higgs field and the dark $U(1)_D$ Higgs field in unitary gauge, respectively, and the Jordan-frame potential $V_{\rm J}(h,\phi)$ is
\begin{align}
	V_{\rm J}(h,\phi) = \frac{1}{4}\lambda_H h^4 + \frac{1}{4}\lambda_D \phi^4 + \frac{1}{4}\lambda_{HD} h^2\phi^2\,.
\end{align}
We note that the VEVs of the SM and dark Higgs fields have been neglected here as they are irrelevant in the inflationary regime.
In this work, we focus on the scenario where the dark Higgs field plays the role of the inflaton and set $h=0$. We call this scenario dark Higgs inflation. We note that the condition for dark Higgs inflation after utilising the minimisation conditions of the inflaton potential for $h=0$ reads \cite{Lebedev:2011aq, Khan:2023uii}
\begin{align}
	\lambda_{HD}\xi_D - 2\lambda_D\xi_H > 0\,.
\end{align}

In order to make a connection between cosmic inflation and low-energy physics of DM, we take into account quantum corrections and consider the renormalisation group (RG)-improved effective action. In the case of dark Higgs inflation, the leading effective action is given by
\begin{align}
	S_{\rm eff} &= \int d^4x \, \sqrt{-g_{\rm J}} \, \bigg\{
	\frac{M_{\rm P}^2}{2}\left[
	1 + \xi_D(t) G_D^2(t) \frac{\phi^2(t)}{M_{\rm P}^2}
	\right]R_{\rm J}
	\nonumber\\&\qquad\qquad\qquad\qquad
	-\frac{1}{2} G_D^2(t) g_{\rm J}^{\mu\nu}\partial_\mu \phi(t) \partial_\nu \phi(t)
	-V_{\rm J,eff}(t)
	\bigg\}\,,
\end{align}
where $t = \ln(\mu/M_Z)$, with $\mu$ and $M_Z$ being, respectively, the renormalisation scale, which we choose to be the inflaton field value in the Jordan frame, and the reference scale of the $Z$ boson mass, and 
\begin{align}
	G_D(t) &= \exp\left(
	-\int^t dt' \frac{\gamma_D}{1+\gamma_D}
	\right)
	\,,\\
	V_{\rm J,eff}(t) &= \frac{1}{4}\lambda_{D,{\rm eff}}(t) G_D^4(t) \phi^4(t)
	\,,
\end{align}
with $\gamma_D=-3g_D^2/(4\pi)^2$ being the anomalous dimension of the dark Higgs field. Here, $\lambda_{D,{\rm eff}}$ is the effective inflaton quartic coupling, which is given by
\begin{align}
	\lambda_{D,{\rm eff}} = \lambda_D + \frac{1}{16\pi^2}\left[
	3g_D^4\left(
	\ln (g_D^2) - \frac{5}{6}
	\right)
	+\lambda_{HD}^2 \left(
	\ln (\lambda_{HD}) - \frac{3}{2}
	\right)
	+9\lambda_D^2 \left(
	\ln (3\lambda_D) - \frac{3}{2}
	\right)
	\right]\,,
\end{align}
where the second term takes into account the one-loop Coleman-Weinberg correction in the $\overline{\rm MS}$ scheme.
The RGEs are summarised in Appendix~\ref{apdx:RGEs}. With the RG-improved effective action, we can move to the Einstein frame, which is a more convenient environment for the inflationary analysis. The resultant Einstein-frame action is given by
\begin{align}
	S_{\rm eff} = \int d^4x \, \sqrt{-g}\left[
	\frac{M_{\rm P}^2}{2}R
	-\frac{1}{2}g^{\mu\nu}\partial_\mu\varphi(t)\partial_\nu\varphi(t)
	-V_{\rm inf}(t)
	\right]\,,
\end{align}
where
\begin{align}
	\left(
	\frac{\partial \varphi}{\partial \phi}
	\right)^2 =
	\frac{G_D^2}{\Omega^2}
	+\frac{3M_{\rm P}^2}{2\Omega^4}\left(
	\frac{d\Omega^2}{d\phi}
	\right)^2
	\,,
\end{align}
with $\Omega^2 = 1+\xi_D G_D^2 \phi^2/M_{\rm P}^2$, and the Einstein-frame effective potential is
\begin{align}\label{eqn:EFeffPot}
	V_{\rm inf}(t) =
	\frac{V_{\rm J,eff}(t)}{\Omega^4}\,.
\end{align}

To proceed the inflationary analysis, we define the so-called slow-roll parameters,
\begin{align}\label{eqn:SRparams}
	\epsilon \equiv \frac{M_{\rm P}^2}{2}\left(\frac{V_{{\rm inf},\varphi}}{V_{\rm inf}}\right)^2
	\,,\quad 
	\eta \equiv M_{\rm P}^2\frac{V_{{\rm inf},\varphi\varphi}}{V_{\rm inf}}
	\,,\quad 
	\kappa^2 \equiv M_{\rm P}^4\frac{V_{{\rm inf},\varphi}V_{{\rm inf},\varphi\varphi\varphi}}{V_{\rm inf}^2}
	\,,
\end{align}
where the comma denotes the derivative with respect to the Einstein-frame field, {\it e.g.}, $V_{{\rm inf},\varphi} \equiv dV_{\rm inf}/d\varphi$, {\it etc.} These slow-roll parameters are then related to the inflationary observables such as the scalar spectral index $n_s$ and the tensor-to-scalar ratio $r$ as follows:
\begin{align}
	n_s &=
	1 - 6\epsilon_* + 2\eta_* 
	- \frac{2}{3}(5+36c)\epsilon_*^2
	+ 2(8c-1)\epsilon_*\eta_* + \frac{2}{3}\eta_*^2
	+\left(\frac{2}{3}-2c\right)\kappa_*^2
	\,,\label{eqn:SSI}\\
	r &=
	16\epsilon_*\left[
	1 + \left(
	4c-\frac{4}{3}
	\right)\epsilon_* + \left(
	\frac{2}{3} - 2c
	\right)\eta_*
	\right]
	\,,\label{eqn:TTSR}
\end{align}
up to the second order in the slow-roll parameters~\cite{Stewart:1993bc,Liddle:1994dx,Leach:2002ar}.
Here, $c=\gamma + \ln 2 - 2$ with $\gamma \approx 0.5772$, and the subscript ``*'' indicates that the quantities are evaluated at the pivot scale of $k=0.05\,{\rm Mpc}^{-1}$. The first slow-roll parameter $\epsilon$ is also related to the magnitude of the curvature power spectrum $A_s$ through
\begin{align}
	A_s = \frac{V_{\rm inf}}{24\pi^2M_{\rm P}^4\epsilon}
	\,,\label{eqn:AsNorm}
\end{align}
which, at the pivot scale, should match $A_s \simeq 2.1\times 10^{-9}$ \cite{Planck:2018jri}.
Classically, this CMB normalisation translates into the relation $\lambda_D/\xi_D^2 \simeq 4.2 \times 10^{-10}$. For the SM Higgs field case where the quartic coupling takes $\lambda_H \simeq 0.12$, the nonminimal coupling is required to be large, $\xi_H \simeq 1.7 \times 10^4$. Such a large value of the nonminimal coupling has raised the concern of unitarity violation \cite{Burgess:2009ea, Barbon:2009ya, Lerner:2009na, Burgess:2010zq, Hertzberg:2010dc, Bezrukov:2010jz}. In the dark Higgs inflation scenario, however, the quartic coupling $\lambda_D$ is a free parameter and may take a small value. Therefore, the nonminimal coupling $\xi_D$ associated with the dark Higgs field does not need to be large.\footnote{
	We note that the unitarity violation scale, which depends on the background inflaton field value, becomes larger than the relevant scale of inflation \cite{Bezrukov:2010jz}. We further note that once the RG running of the coupling parameters is taken into account, the SM Higgs quartic coupling may become extremely small, making the nonminimal coupling small as well \cite{Bezrukov:2014bra, Hamada:2014iga, Kim:2014kok, Hamada:2014wna}. However, it requires a fine-tuning of the parameters, and the tensor-to-scalar ratio $r$ becomes large.
}
In our numerical analysis, we fix the nonminimal coupling $\xi_D$ using the CMB normalisation condition; the $\xi_D$ parameter is thus not a free parameter. The constraints on $n_s$ and $r$ are $0.958 \leq n_s \leq 0.975$ (95\% C.L.) and $r \leq 0.036$ (95\% C.L.) \cite{Planck:2018jri, BICEP:2021xfz}. Recently, the Atacama Cosmology Telescope (ACT) collaboration has reported the combined analysis of their data, the Planck data, and the baryon acoustic oscillation data, favouring a higher value of the spectral index $n_s = 0.974 \pm 0.003$ \cite{ACT:2025fju, ACT:2025tim}.

We closely follow Ref.~\cite{Khan:2023uii} to compute the inflationary observables $n_s$ and $r$; see also Refs.~\cite{Lerner:2009xg, Lerner:2011ge, Kim:2014kok}. For completeness, we briefly outline our methodology. For a given set of BSM parameters, $\{M_{h_2}, M_{W_D}, g_D, \sin\alpha\}$, we find the values of all the model parameters at the initial scale of $t=0$. We then choose a value for the nonminimal coupling parameter $\xi_D$ with $\xi_H=0$ at $t=0$ and solve the RGEs given in Appendix~\ref{apdx:RGEs} to a high-energy scale. We construct the RG-improved effective potential \eqref{eqn:EFeffPot} and compute the slow-roll parameters \eqref{eqn:SRparams}. The end of inflation is then found by imposing $\epsilon=1$. To find the inflaton field value at the pivot scale $\varphi_*$, we impose $N_*$ number of $e$-folds which is given by our choice of the reheating temperature $T_{\rm reh}$ as we discuss below. The magnitude of the curvature power spectrum is evaluated at the pivot scale through Eq.~\eqref{eqn:AsNorm}. If it does not match $2.1 \times 10^{-9}$, we vary the nonminimal coupling parameter $\xi_D$ and repeat the procedure until we find the value that satisfies the CMB normalisation. Finally, we compute the spectral index $n_s$ and the tensor-to-scalar ratio $r$ using Eqs.~\eqref{eqn:SSI} and \eqref{eqn:TTSR}.

Let us now discuss how the number of $e$-folds $N_*$ is determined. The number of $e$-folds is given, in the slow-roll approximation, by
\begin{align}
	N_* = -\frac{1}{M_{\rm P}^2}\int_{\varphi_*}^{\varphi_e} \frac{V_{\rm inf}}{V_{{\rm inf},\varphi}} d\varphi
	\,,\label{eqn:efolds}
\end{align}
where the subscript $e$ denotes the end of inflation.
This quantity $N_*$ is related to the reheating temperature $T_{\rm reh}$ through \cite{Cook:2015vqa}
\begin{align}
	T_{\rm reh} = \left(
	\frac{43}{11g_{s,{\rm reh}}}
	\right)^{1/3}\left(
	\frac{a_0T_0}{k_*}
	\right)H_* e^{-N_*}e^{-N_{\rm reh}}
	\,,
\end{align}
where $g_{s,{\rm reh}}$ is the effective number of relativistic degrees of freedom at reheating, $T_0$ is the present-day temperature, $k_* = 0.05\,{\rm Mpc}^{-1}$ is the pivot scale, $H_*$ is the Hubble parameter at the horizon exit of the pivot scale, and $N_{\rm reh}$ is the number of $e$-folds during reheating. We may convert the relation to find
\begin{align}
	N_* = \frac{1}{3}\ln\left(
	\frac{43}{11g_{s,{\rm reh}}}
	\right)
	+\ln\left(
	\frac{a_0T_0}{k_*}
	\right)
	+\ln\left(
	\frac{H_*}{T_{\rm reh}}
	\right)
	-N_{\rm reh}
	\,.
\end{align}
In our scenario, after the end of inflation, the Einstein-frame potential takes the form (see also Refs.~\cite{Bezrukov:2008ut, Gong:2015qha})
\begin{align}
	V_{\rm inf} \approx \begin{cases}
		\frac{\lambda_D M_{\rm P}^2}{6\xi_D^2}\varphi^2 & \text{ for } \sqrt{2/3}M_{\rm P}/\xi_D < \varphi \ll M_{\rm P} \,,\\
		\frac{1}{4}\lambda_D \varphi^4 & \text{ for } v_D \ll \varphi < \sqrt{2/3}M_{\rm P}/\xi_D \,,\\
		\lambda_D v_D^2 \varphi^2 & \text{ for } \varphi \ll v_D \,,
	\end{cases}
	\label{inflaton-potential}
\end{align}
for $\xi_D \gtrsim 1.46$ \cite{Gong:2015qha}.
We see that the potential form changes during the reheating phase, from quadratic to quartic and back to quadratic. We call these regimes Stage 1, Stage 2, and Stage 3, respectively. Equivalently, the equation of state $w$ changes during the course of reheating.
Thus, we may generically write
\begin{align}
	N_{\rm reh} = N_{\rm reh}^{(1)} + N_{\rm reh}^{(2)} + N_{\rm reh}^{(3)}\,,
\end{align}
where
\begin{gather}
	N_{\rm reh}^{(1)} = \frac{1}{3(1+w_1)}\ln\left(
	\frac{\rho_{\rm end}}{\rho_1}
	\right)\,,\;
	N_{\rm reh}^{(2)} = \frac{1}{3(1+w_2)}\ln\left(
	\frac{\rho_1}{\rho_2}
	\right)\,,\;
	N_{\rm reh}^{(3)} = \frac{1}{3(1+w_3)}\ln\left(
	\frac{\rho_2}{\rho_{\rm reh}}
	\right)\,,
\end{gather}
with $\rho_1$ ($\rho_2$) being the inflaton energy density at the end of the first (second) stage of reheating, $w_1 = 0$, $w_2 = 1/3$, $w_3 = 0$, and
\begin{align}
	\rho_{\rm reh} = \frac{\pi^2}{30}g_{\rho,{\rm reh}}T_{\rm reh}^4\,.
\end{align}
We note that if reheating ends during Stage 2, $N_{\rm reh}^{(3)}$ needs to be set to zero together with $\rho_2 = \rho_{\rm reh}$.
In principle, the value $H_*$ depends on $N_*$. It is important to note that the model parameters are running under the RGEs. In particular, the model parameters at the end of inflation differ from those at the horizon exit of the pivot scale $k_*$. Therefore, it is difficult to express $N_*$ in terms of the model parameters analytically. Instead, we numerically iterate the above relation to find $N_*$ that gives rise to the desired value of $T_{\rm reh}$.
Currently, only the quadratic potential is implemented in {\tt micrOMEGAs}, {\it i.e.}, only $w = 0$ is available. This is not always the case in dark Higgs inflation as we have discussed above. To be more precise, one needs to modify the underlying routine of {\tt micrOMEGAs}, which goes beyond the scope of the current work. We instead keep the $w=0$ assumption for the DM study, while for the inflationary analysis, we take into account the variation of the equation of state. The modification of {\tt micrOMEGAs} for the exact implementation of the DM analysis will be reported elsewhere in the future. We note that the assumption of $w=0$ for the DM analysis could be traded for the uncertainty in the physics of reheating; the uncertainty in $N_{\rm reh}$, for instance, does not drastically change $n_s$ or $r$ as is also shown in, {\it e.g.}, Ref.~\cite{Gong:2015qha}.
It is interesting to note that the duration of the first reheating stage is given by $N_{\rm reh}^{(1)} \approx (1/3)\ln (0.675\xi_D^2)$. Therefore, we conclude that the first stage is rather short; we have $N_{\rm reh}^{(1)} = 6$ even if a large value of $\xi_D = 10^4$ is considered. The inflaton then enters the regime of $\varphi < \sqrt{2/3}M_{\rm P}/\xi_D$.

Before we end this section, we comment on the reheating temperature $T_{\rm reh}$ in the dark Higgs inflationary scenario. We recall that the discussion of DM in section~\ref{sec:dm} involves a low reheating temperature. A low reheating temperature is difficult to achieve in the SM Higgs inflationary scenario where $T_{\rm reh} \simeq 10^{13-15}\,{\rm GeV}$ is typically estimated \cite{Bezrukov:2007ep, Bezrukov:2008ut}. This is, in essence, due to the nature of strong interactions between the SM Higgs field and the rest of the SM fields. In our dark Higgs inflationary scenario, however, it is not necessarily the case. The interaction between the dark Higgs field, which is the inflaton, and the SM thermal bath is through the Higgs-portal coupling $\lambda_{HD}$. The portal coupling $\lambda_{HD}$ is a free parameter and could easily be small, lowering the reheating temperature. In fact, the portal coupling $\lambda_{HD}$ is related to the mixing angle between the SM Higgs and dark Higgs bosons, and it should be small to be consistent with the Higgs signal strength measurements at the LHC.

\subsection{Reheating dynamics}
\label{subsec:reheating}
\begin{table}[]
	\centering
	\begin{tabular}{|c|c|c|}
		\hline
		Reheating Stage & Decay & Scattering \\ \hline
		Stage 2 ($a_{1} < a_{\rm reh} < a_{2}$) & $\lambda_{D} > 2 \lambda_{HD}/3$  & $\lambda_{HD}/6<\lambda_{D} < 2 \lambda_{HD}/3$ \\ \hline
		Stage 3 ($a_{\rm reh} > a_{2}$) & $M_{h_2} > 2 M_{h_1}$ & $M_{h_1} < M_{h_2} < 2 M_{h_1}$ \\
		\hline
	\end{tabular}
	\caption{Conditions for achieving decay- and scattering-dominated reheating. Here, $a$ is the scale factor with $a_1$ and $a_2$ being the scale factors at the start of the second reheating stage, {\it i.e.}, $\varphi = \sqrt{2/3} M_{\rm P} / \xi_D$, and when the inflaton field value hits the dark Higgs VEV $v_D$. During Stage 2, the inflaton value is still larger than $v_D$, and the effective SM and dark Higgs masses are inflaton field-dependent. On the other hand, in the $a > a_2$ regime, the effective masses get reduced to the low-energy physical masses $M_{h_1}$ and $M_{h_2}$.}
	\label{tab:conditions}
\end{table}

For the reheating dynamics in dark Higgs inflation, one may take the perturbative decay reheating scenario where the inflaton decays into a pair of the SM Higgs through the interaction term
\begin{align}
	\mathcal{L}_{\phi h^2} = \frac{1}{2}\lambda_{HD}v_D\phi h^2\,,
\end{align}
as long as the inflaton is heavier than the twice of the SM Higgs mass. One may also consider the scattering process through the interaction term
\begin{align}
	\mathcal{L}_{\phi^2 h^2} = \frac{1}{4}\lambda_{HD} \phi^2 h^2\,.
\end{align}
The SM Higgs mass and the dark Higgs inflaton mass here should be regarded as the effective masses during reheating, and they are respectively given by
\begin{align}
	M_{h,{\rm eff}} \approx \sqrt{\frac{\lambda_{HD}}{2}}\varphi
	\,,\qquad
	M_{\varphi,{\rm eff}} \approx \sqrt{3\lambda_D}\varphi
	\,,
\end{align}
during the initial period of the second stage when the inflaton field value is smaller than $\sqrt{2/3}M_{\rm P}/\xi_D$ but still larger than the dark Higgs VEV $v_D$, {\it i.e.}, $v_D < \varphi < \sqrt{2/3} M_P / \xi$. On the other hand, when $\varphi$ becomes smaller than $v_D$, the effective masses reduce to the low-energy physical masses, $M_{h,{\rm eff}} = M_{h_1}$ and $M_{\varphi,{\rm eff}} = M_{h_2}$. As long as the produced particles are lighter than the inflaton, $M_{\varphi,{\rm eff}} > 2 M_{h,{\rm eff}}$ (or $M_{\varphi,{\rm eff}} > M_{h,{\rm eff}}$), one may consider the perturbative decay (or scattering) for reheating; see Table~\ref{tab:conditions}. We stress again that the first reheating stage is short, and the inflaton quickly enters the second reheating stage $\varphi < \sqrt{2/3}M_{\rm P}/\xi_D$. The production of the radiation during the first stage is thus negligible, and we may safely consider the evolution of the radiation from the second stage.

In Ref.~\cite{Garcia:2020wiy}, reheating dynamics is studied when the inflaton field $\varphi$ is oscillating around the potential $V(\varphi) \sim \varphi^k$ after the end of inflation. In our dark Higgs inflation scenario, the value of $k$ is not a fixed constant but rather changes depending on the field value. Closely following the discussion presented in Ref.~\cite{Garcia:2020wiy}, we track the evolutions of the inflaton energy density and the radiation energy density by taking into account the fact that the inflaton potential form changes over the course of reheating. During the second and third reheating stages after the short first stage, the potential form changes once; from the quartic form when the inflaton value is larger than the dark Higgs VEV, $\varphi > v_D$, to the quadratic form when $\varphi < v_D$. The inflaton energy density during these stages evolves as
\begin{align}
	\rho_{\rm inf}(a) =
	\begin{cases}
		\rho_{1}\left(\frac{a}{a_{1}}\right)^{-4}
		\,,&\text{ for $a_{1} < a < a_2$}\,,\\
		\rho_{2}\left(\frac{a}{a_{2}}\right)^{-3}
		\,,&\text{ for $a_{2} < a$}\,,
	\end{cases}
\end{align}
where $a_1$ is the scale factor at the end of the first reheating stage, and $a_2$ is the scale factor at the end of the second reheating stage. Here, $\rho_1$ and $\rho_2$ are the inflaton energy densities at $a_1$ and $a_2$, respectively, and we approximate them as the potential energy at $a_1$ and $a_2$.
On the other hand, the evolution of the radiation energy density is governed by
\begin{align}
	\frac{1}{a^4}\frac{d}{da}\left(\rho_R a^4\right) = \frac{2k}{k+2}\frac{\gamma_{\rm inf}}{aH}\frac{\rho_{\rm inf}^{l+1}}{M_{\rm P}^{4l}}\,,
\end{align}
where $l$ is defined through the form of the interaction rate \cite{Garcia:2020wiy},
\begin{align}
	\Gamma_{\rm inf} = \gamma_{\rm inf}\left(\frac{\rho_{\rm inf}}{M_{\rm P}^4}\right)^l\,.
\end{align}

Let us consider the decay process. In this case, assuming that $\rho_R$ is negligible initially, we find
\begin{align}
	\rho_R \approx
	\frac{\sqrt{3} \lambda_{HD}^2 v_D^2 M_{\rm P} \rho_1^{1/4}}{48 \sqrt{\pi} \lambda_D^{1/4}} \frac{\Gamma(3/4)}{\Gamma(1/4)}
	\left(
	\frac{a_1}{a}
	\right)\left[
	1 - \left(
	\frac{a_1}{a}
	\right)^3
	\right]
	\,,
\end{align}
in the $a_1 < a < a_2$ region, where we have neglected the kinematic factor, which does not change the conclusion when the SM Higgs mass is much less than or similar to the inflaton mass \cite{Garcia:2020wiy}.
In the $a_2 < a$ region, we obtain
\begin{align}
	\rho_R \approx
	\frac{\sqrt{6} \lambda_{HD}^2 v_D M_{\rm P} \sqrt{\rho_2}}{160 \pi \sqrt{\lambda_D}}
	\left(
	\frac{a_2}{a}
	\right)^{3/2}\left[ 
	1 - \left(
	\frac{a_2}{a}
	\right)^{5/2}
	\right]
	+\rho_R(a_2)\left(
	\frac{a_2}{a}
	\right)^4
	\,,
\end{align}
where
\begin{align}
	\rho_R(a_2) \approx
	\frac{\sqrt{3} \lambda_{HD}^2 v_D^2 M_{\rm P} \rho_1^{1/4}}{48 \sqrt{\pi} \lambda_D^{1/4}} \frac{\Gamma(3/4)}{\Gamma(1/4)}
	\left(
	\frac{a_1}{a_2}
	\right)
	\,,
\end{align}
with $a_2 \gg a_1$ assumed.
At the end of reheating, $\rho_{\rm inf}(a_{\rm reh}) = \rho_R(a_{\rm reh})$. If $a_{\rm reh} > a_2$, we have
\begin{align}
	\rho_R(a_{\rm reh}) &\approx 
	\frac{\sqrt{6} \lambda_{HD}^2 v_D M_{\rm P} \sqrt{\rho_2}}{160 \pi \sqrt{\lambda_D}}
	\left(
	\frac{a_2}{a_{\rm reh}}
	\right)^{3/2}
	+\frac{\sqrt{3} \lambda_{HD}^2 v_D^2 M_{\rm P} \rho_1^{1/4}}{48 \sqrt{\pi} \lambda_D^{1/4}} \frac{\Gamma(3/4)}{\Gamma(1/4)}
	\left(
	\frac{a_1}{a_2}
	\right)\left(
	\frac{a_2}{a_{\rm reh}}
	\right)^4
	\,,
\end{align}
and
\begin{align}
	\rho_{\rm inf}(a_{\rm reh}) = \rho_{2}\left(\frac{a_{\rm reh}}{a_{2}}\right)^{-3}\,.
\end{align}
Here, we may express $a_2 = a_1 (\rho_1 / \rho_2)^{1/4}$.
On the other hand, if $a_1 < a_{\rm reh} < a_2$, we would have
\begin{align}
	\rho_R(a_{\rm reh}) \approx 
	\frac{\sqrt{3} \lambda_{HD}^2 v_D^2 M_{\rm P} \rho_1^{1/4}}{48 \sqrt{\pi} \lambda_D^{1/4}} \frac{\Gamma(3/4)}{\Gamma(1/4)}
	\left(
	\frac{a_1}{a_{\rm reh}}
	\right)
	\,,
\end{align}
and
\begin{align}
	\rho_{\rm inf}(a_{\rm reh}) = \rho_1 \left( \frac{a_{\rm reh}}{a} \right)^{-4}\,.
\end{align}
Numerically finding $a_{\rm reh}/a_1$ or $a_{\rm reh}/a_2$ from $\rho_{\rm inf}(a_{\rm reh}) = \rho_R(a_{\rm reh})$ and using
\begin{align}
	T_{\rm reh} = \left[
	\frac{30}{\pi^2 g_{*,{\rm reh}}}\rho_R(a_{\rm reh})
	\right]^{1/4}\,,
\end{align}
we can read the reheating temperature.

\begin{figure}[t!]
	\centering
	\includegraphics[width=0.48\linewidth]{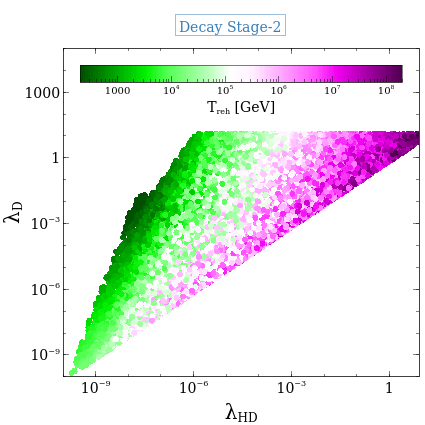}
	\includegraphics[width=0.48\linewidth]{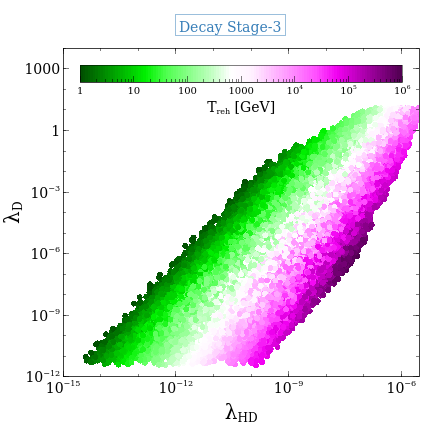}
	\caption{Reheating temperature $T_{\rm reh}$ in the $\lambda_{HD}$--$\lambda_{D}$ plane for the decay process scenario. The left panel corresponds to the case where the end of reheating occurs in the second stage, while the right panel is when reheating ends during the third stage. We observe that the reheating temperature decreases as the coupling $\lambda_{HD}$ decreases, as anticipated, and the reheating temperature can be as small as $T_{\rm reh} = 1$ GeV. The perturbativity conditions, $\{\lambda_H, \lambda_D, \lambda_{HD}\} < 4\pi$, are imposed, and we have required that the dark Higgs decays before the BBN.}
	\label{fig:Treh-decay}
\end{figure}
The resultant reheating temperature for the decay process is presented in Fig.~\ref{fig:Treh-decay} for the following parameter ranges:
\begin{align}
	10^{2} \leq M_{h_2} \, [{\rm GeV}] \leq 10^{5}
	\,,\quad
	10^{-9} \leq \sin \alpha \leq 0.23
	\,,\quad
	10^{3} \leq v_{D} \, [{\rm GeV}] \leq 10^{8},
	\label{eqn:range-parameter-Treh}
\end{align}
where the upper bound on $\sin\alpha$ comes from the Higgs precision data.
As we have outlined in Table~\ref{tab:conditions}, we have imposed $\lambda_D > 2\lambda_{HD}/3$ ($M_{h_2} > 2 M_{h_1}$) if the reheating ends in the $a_1 < a < a_2$ ($a > a_2$) regime.
The left panel of Fig.~\ref{fig:Treh-decay} shows the case when reheating ends during Stage 2, {\it i.e.}, $a_1 < a_{\rm reh} < a_2$, whereas the right panel corresponds to the case when the end of reheating occurs during Stage 3, {\it i.e.}, $a_{\rm reh} > a_2$. In addition to the kinematic conditions for the decay, we have imposed the perturbativity conditions for the quartic couplings, namely $\{\lambda_H, \lambda_D, \lambda_{HD}\} < 4\pi$ and required that the dark Higgs decays before the Big Bang Nucleosynthesis (BBN) to keep the BBN physics intact. We see that the reheating temperature can be as low as $T_{\rm reh} = 1$ GeV and that the coupling $\lambda_{HD}$ needs to be small or, equivalently, the Higgs mixing angle $\alpha$ needs to be small, in order to achieve a low reheating temperature as well as to be consistent with the Higgs properties measured at the LHC.

Let us now consider the scattering process $\phi \phi \rightarrow h h$ by which the SM Higgs bosons, and subsequently the SM bath particles, are produced. Following the same procedure as before, we find
\begin{align}
	\rho_R \approx 
	\frac{\sqrt{3} \lambda_{HD}^2 M_{\rm P} \rho_1^{3/4}}{64\sqrt{\pi} \lambda_D^{3/4}} \frac{\Gamma(3/4)}{\Gamma(1/4)}
	\left(
	\frac{a_1}{a}
	\right)^3\left[
	1 - \frac{a_1}{a}
	\right]
	\,,
\end{align}
in the $a_1 < a < a_2$ regime, and
\begin{align}
	\rho_R \approx 
	\frac{\sqrt{3} \lambda_{HD}^2 M_{\rm P} \rho_2^{3/2}}{128 \sqrt{2} \pi \lambda_D^{3/2} v_D^3}
	\left(
	\frac{a_2}{a}
	\right)^4\left[ 
	1 - \left(
	\frac{a_2}{a}
	\right)^{1/2}
	\right] + \rho_R(a_2)\left(
	\frac{a_2}{a}
	\right)^4
	\,,
\end{align}
in the $a_2 < a$ regime, where
\begin{align}
	\rho_R(a_2) \approx \frac{\sqrt{3} \lambda_{HD}^2 M_{\rm P} \rho_1^{3/4}}{64\sqrt{\pi} \lambda_D^{3/4}} \frac{\Gamma(3/4)}{\Gamma(1/4)}
	\left(
	\frac{a_1}{a_2}
	\right)^3\,,
\end{align}
with $a_2 \gg a_1$ assumed.
We again numerically find $a_{\rm reh}/a_1$ or $a_{\rm reh}/a_2$ from $\rho_{\rm inf}(a_{\rm reh}) = \rho_R(a_{\rm reh})$, and then the reheating temperature is obtained by using
\begin{align}
	T_{\rm reh} = \left[
	\frac{30}{\pi^2 g_{*,{\rm reh}}}\rho_R(a_{\rm reh})
	\right]^{1/4}\,.
\end{align}
\begin{figure}
	\centering
	\includegraphics[width=0.48\linewidth]{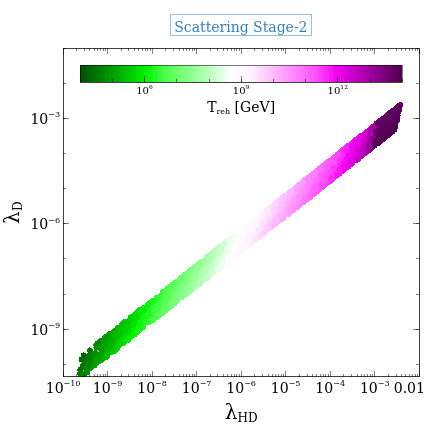}
	\includegraphics[width=0.48\linewidth]{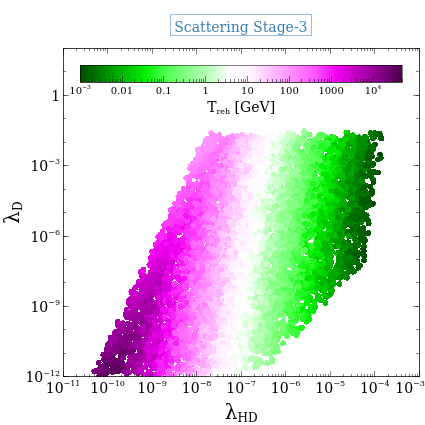}
	\caption{Reheating temperature $T_{\rm reh}$ in the $\lambda_{HD}$--$\lambda_{D}$ plane for the scattering process scenario. The left (right) panel corresponds to the case where reheating ends during the second (third) stage. In the case of the Stage 2 reheating, the reheating temperature grows as $\lambda_{HD}$, while when the end of reheating occurs during Stage 3, it shows an opposite behaviour, and the reheating temperature increases as $\lambda_{HD}$ decreases. We observe that the reheating temperature can be as low as $T_{\rm reh} \simeq 10^4$ GeV in the case of the Stage 2 reheating. In the case of the Stage 3 reheating, the reheating temperature may become very small, even close to the BBN bound $T_{\rm reh} \simeq 1$ MeV, with a relatively large value of $\lambda_{HD} \simeq 10^{-4}$. The perturbativity conditions, $\{\lambda_H, \lambda_D, \lambda_{HD}\} < 4\pi$, are imposed, and we have required that the dark Higgs decays before the BBN. Furthermore, $T_{\rm reh} > 1$ MeV is considered.}
	\label{fig:Treh-scattering}
\end{figure}

In Fig.~\ref{fig:Treh-scattering}, we present the reheating temperature for the scattering process with the same parameter ranges as in the decay process case \eqref{eqn:range-parameter-Treh}. The left (right) panel of Fig.~\ref{fig:Treh-scattering} shows the case where the end of reheating happens during the second (third) stage, {\it i.e.}, $a_1 < a_{\rm reh} < a_2$ ($a > a_2$). In addition to the kinematic constraints displayed in Table~\ref{tab:conditions}, we have imposed the perturbativity conditions for the quartic couplings, $\{\lambda_H, \lambda_D, \lambda_{HD}\} < 4\pi$ and required that the dark Higgs decays before the BBN. Furthermore, the points shown in Fig.~\ref{fig:Treh-scattering} satisfy $T_{\rm reh} > 1$ MeV, which we take to be the lower bound on the reheating temperature; see, for instance, Refs.~\cite{Kawasaki:1999na, Kawasaki:2000en, Hasegawa:2019jsa} for an in-depth analysis. In the case of the Stage 2 reheating, we observe a narrow allowed parameter region because of the kinematic conditions imposed on the quartic couplings $\lambda_{HD}$ and $\lambda_{D}$. The reheating temperature in this case can be as low as $T_{\rm reh} \simeq 10^{4}$ GeV. In the case of the Stage 3 reheating, we can achieve the reheating temperature at the MeV scale. 

We thus conclude that a low reheating temperature could be achieved from both the decay and scattering cases in the model under consideration if we take the dark Higgs inflationary scenario. This result supports our consideration of DM phenomenology in the context of the low reheating temperature discussed in section~\ref{sec:dm}. 

In addition to the SM bath production, it is also important to consider the production of DPDM from the inflaton. This production channel should be suppressed in our case, for the FIMP-type DM, in particular. The effective DPDM mass in the Einstein frame is given by
\begin{align}
	M_{W_D,{\rm eff}} \approx g_D \varphi \,,
\end{align}
where we have again focused on the second stage of reheating. In the third stage, it reduces to $M_{W_D} = g_D v_D$. Recalling the inflaton mass, we see that if $g_D > \sqrt{3\lambda_D}$ for $\varphi < \sqrt{2/3}M_{\rm P}/\xi_D$, we have $M_{\varphi,{\rm eff}} < M_{W_D,{\rm eff}}$. The DPDM production directly from the inflaton is thus kinetically forbidden in this case. The very same condition of $g_D > \sqrt{3\lambda_D}$ also forbids the direct DPDM production from the inflaton during the first stage of reheating, where $\sqrt{2/3}M_{\rm P}/\xi_D < \varphi \ll M_{\rm P}$, as long as $\xi_D \gtrsim 1$, as the effective masses are given by $M_{W_D,{\rm eff}} \approx (2/3)^{1/4} g_D \sqrt{M_{\rm P}\varphi/\xi_D}$ and $M_{\varphi,{\rm eff}} \approx \sqrt{\lambda_D}M_{\rm P}/(\sqrt{3}\xi_D)$. During the parameter scan reported in the next section, we impose the condition $g_D \gtrsim \sqrt{\lambda_D}$ not to jeopardise the FIMP-type DM scenario.

\section{Results}
\label{sec:results}
We now present results of the parameter scan and identify the allowed parameter space. Based on the discussions of sections~\ref{sec:dm} and \ref{sec:inflation}, we vary the model parameters in the following ranges:
\begin{gather}
	10^{2} \leq M_{h_2} \, [{\rm GeV}] \leq 10^{7}
	\,,\quad 
	10^{2} \leq M_{W_D} \, [{\rm GeV}] \leq 10^{7}
	\,,\nonumber\\
	10^{-7} \leq g_{D} \leq 1
	\,,\quad 
	10^{-8} \leq \sin\alpha \leq 0.23
	\,,\quad
	10^{-3} \leq \frac{T_{\rm reh}}{M_{W_D}} \leq 1
	\,.\label{eqn:ParamScanRegion}
\end{gather}
We stress again that the nonminimal coupling parameter $\xi_D$ is determined by the CMB normalisation, $A_s \simeq 2.1 \times 10^{-9}$, and thus, it is not a free parameter.

\begin{figure}[t!]
	\centering
	\includegraphics[angle=0,height=7cm,width=7.5cm]{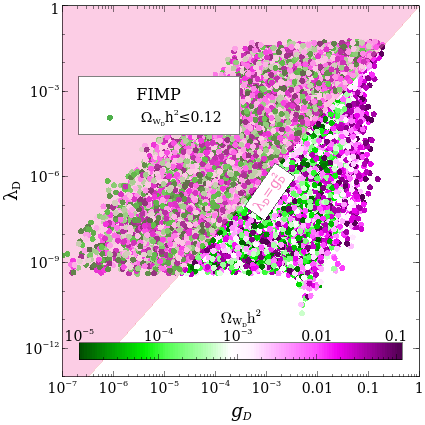}
	\includegraphics[angle=0,height=7cm,width=7.5cm]{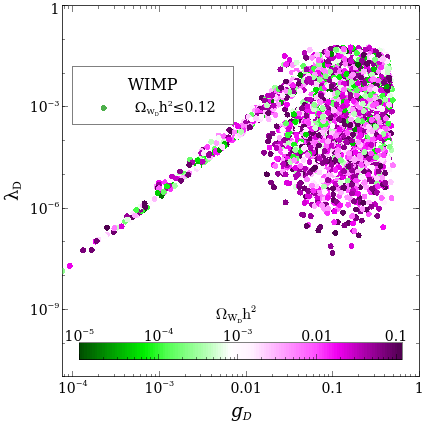}
	\caption{DM relic density in the $g_{D}$--$\lambda_{D}$ plane for the FIMP-type (left) and WIMP-type (right) DM candidates. The red-shaded region in the left panel corresponds to $g_{D} < \sqrt{\lambda_{D}}$ which kinematically allows the direct decay of the inflaton to DM.} 
	\label{fig:scatter-plot-1}
\end{figure}

In Fig.~\ref{fig:scatter-plot-1}, we show the DM relic density $\Omega_{W_D}h^2$ in the $g_{D}$--$\lambda_{D}$ plane. The left panel corresponds to the FIMP-type DM scenario, where DM is produced by the freeze-in mechanism as defined in section~\ref{sec:dm}. The red-shaded region is where $g_D < \sqrt{\lambda_D}$ which allows the direct decay of the inflaton to DM, and the white region kinematically forbids such a direct decay. We observe that a wide range of the parameter space is allowed.
The right panel, on the other hand, corresponds to the WIMP-type DM scenario. We can see here that a larger value of the dark gauge coupling $g_D$ is generally required for the DM production through the freeze-out mechanism. A very sharp region allowed in the $g_{D}$--$\lambda_{D}$ plane corresponds to the dark Higgs resonance region. In the WIMP case, the direct decay of the inflaton is not problematic as the produced DM from such a decay would reach the thermal equilibrium.

\begin{figure}[t!]
	\centering
	\includegraphics[angle=0,height=7cm,width=7.5cm]{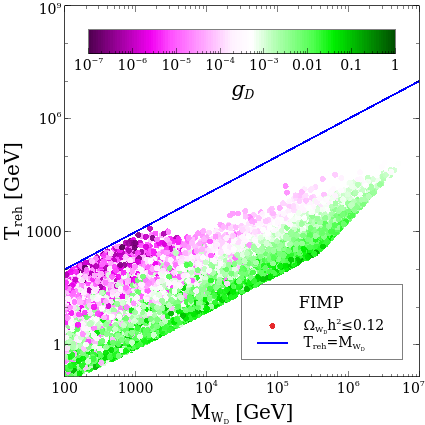}
	\includegraphics[angle=0,height=7cm,width=7.5cm]{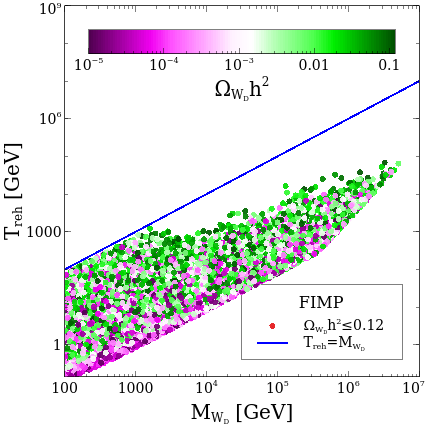}\\
	\includegraphics[angle=0,height=7cm,width=7.5cm]{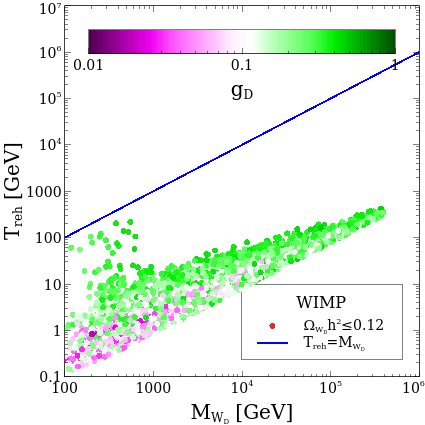}
	\includegraphics[angle=0,height=7cm,width=7.5cm]{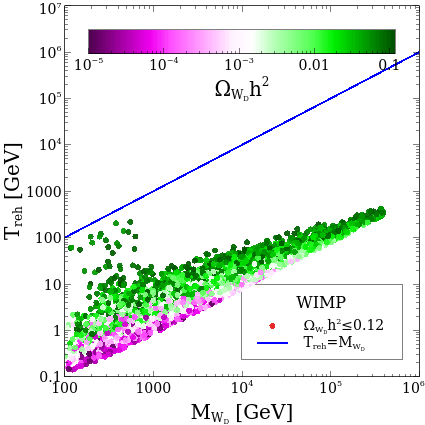}
	\caption{Allowed parameter space in the $M_{W_D}$--$T_{\rm reh}$ plane. The colour variation represents the dark gauge coupling $g_D$ in the left panel and the DM relic density $\Omega_{W_D}h^{2}$ in the right panel. The top (bottom) panel is for the FIMP-type (WIMP-type) DM scenario.}
	\label{fig:scatter-plot-3}
\end{figure}

In Fig.~\ref{fig:scatter-plot-3}, we present the allowed parameter space in the $M_{W_D}$--$T_{\rm reh}$ plane. The upper panel is for the FIMP-type DM, and the lower panel is for the WIMP-type DM. The colour variation indicates values of the dark gauge coupling $g_{D}$ in the left panel, while in the right panel, it indicates the DM relic density. In the top-left plot, we see that there is no point allowed for $T_{\rm reh} > M_{W_D}$ where the DM gets overproduced. For a fixed value of the DM mass $M_{W_D}$, as the reheating temperature increases, we need lower values of the dark gauge coupling $g_D$ to obtain the DM relic density. This is because the amount of dilution due to the entropy production reduces as $T_{\rm reh}$ increases, and hence, we need a lower value of $g_D$ to obtain the DM density as it is proportional to the dark gauge coupling. Moreover, we have fewer allowed points for $M_{W_D} > 10^{5}$ GeV. It is because DM is overproduced in those regions due to less dilution from the entropy production, moving towards the standard freeze-in mechanism, where we need a lower dark gauge coupling. In the top-right plot, the DM relic density is presented. For a fixed value of the DM mass $M_{W_D}$, the DM relic density shows an increasing tendency as we increase $T_{\rm reh}$, although overlapping points also exist due to the variation of the dark gauge coupling.
The bottom-left plot shows the variation of the dark gauge coupling $g_D$ for the WIMP-type DM produced by the freeze-out mechanism. In this case, for lower values of $T_{\rm reh}$, we need lower values of $g_D$ to obtain the DM relic density in the allowed range, an opposite behaviour to the FIMP-type DM scenario. This can be understood by noting that $\Omega^{\rm WIMP}_{W_D} h^{2} \propto 1/\langle \sigma v \rangle_{W_D W_D \rightarrow h_2 h_2}$. In the bottom-right plot, the DM relic density is presented for the WIMP-type DM. We clearly see that lower values of $T_{\rm reh}$ lead to lower values of the DM relic abundance which stems from entropy dilution. This behaviour is consistent with the FIMP-type DM scenario because entropy dilution is related to the reheating temperature, not the DM production mechanism. For the WIMP-type DM produced by the freeze-out mechanism, we note that the $M_{W_D} > 10^{5}$ GeV region is still allowed due to entropy dilution.

\begin{figure}[t!]
	\centering
	\includegraphics[angle=0,height=7cm,width=7.5cm]{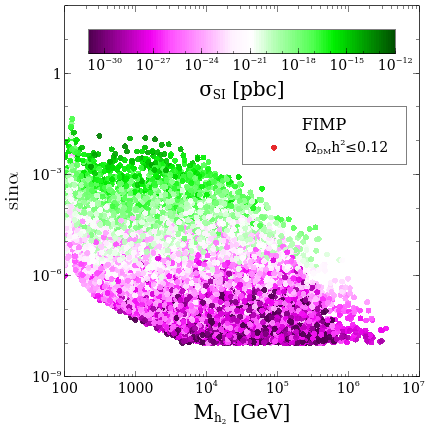}
	\includegraphics[angle=0,height=7cm,width=7.5cm]{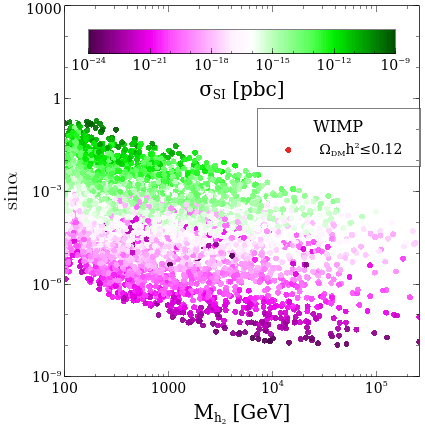}
	\caption{DM direct detection cross-section in the $M_{h_2}$--$\sin\alpha$ plane for the FIMP-type (left) and WIMP-type (right) DM scenarios.}
	\label{fig:scatter-plot-4}
\end{figure}

The DM direct detection cross-section $\sigma_{\rm SI}$, Eq.~\eqref{eqn:dd-cs}, is shown in Fig.~\ref{fig:scatter-plot-4} in the $M_{h_2}$--$\sin\alpha$ plane. The left panel is for the FIMP-type DM scenario, and the right panel is for the WIMP-type DM scenario. In both cases, we observe that a lower Higgs mixing angle $\sin\alpha$ is required for a larger dark Higgs mass $M_{h_2}$. This behaviour is tied to the condition $\lambda_{D} > \lambda_{HD}$, allowing for the inflaton, which in our case is the dark Higgs field, to decay to the SM Higgs and produce the radiation bath. The slight gap near $M_{h_2} \simeq 125$ GeV is due to the cancellation in the direct detection cross-section. The general trend in both the freeze-in case and the freeze-out case is that the direct detection cross-section varies with the square of the Higgs mixing angle, {\it i.e.}, $\sigma_{\rm SI} \propto \sin^2\alpha$. Additionally, it does not change significantly with the change in the dark Higgs mass because the $h_2$ contribution goes down with the increase in its mass.

\begin{figure}[t!]
	\centering
	\includegraphics[angle=0,height=7cm,width=7.5cm]{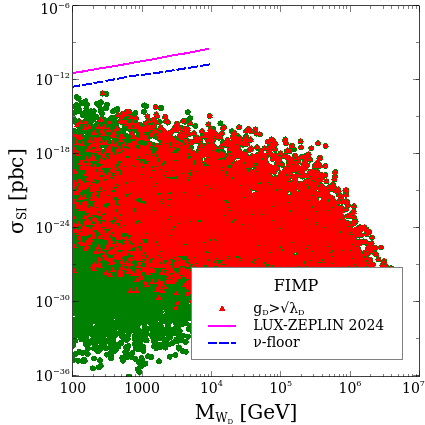}
	\includegraphics[angle=0,height=7cm,width=7.5cm]{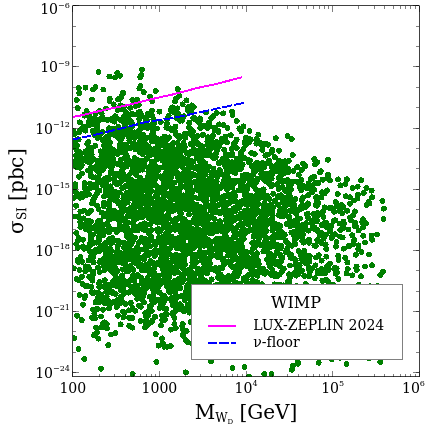}
	\caption{DM direct detection cross-section $\sigma_{\rm SI}$ in terms of the DM mass $M_{W_D}$. The left (right) panel corresponds to the FIMP-type (WIMP-type) DM scenario. The points are obtained after imposing the upper bound on the DM relic density. The green solid lines represent the LUX-ZEPLIN 2024 bound, while the blue dashed lines indicate the neutrino floor bound. The green points in the left panel are those that allow the direct decay of the inflaton to DM, satisfying $g_{D} > \sqrt{\lambda_{D}}$.} 
	\label{fig:scatter-plot-5}
\end{figure}

Figure~\ref{fig:scatter-plot-5} presents the DM direct detection cross-section $\sigma_{\rm SI}$ in terms of the DM mass $M_{W_D}$ after imposing the aforementioned constraints. The left panel is for the FIMP-type DM, whereas the right panel is for the WIMP-type DM. In the left panel, the green points correspond to the parameter space where $g_{D} > \sqrt{\lambda_{D}}$ is satisfied, kinematically forbidding the direct decay of the inflaton to DPDM. For the FIMP-type DM scenario, all the allowed points sit below the neutrino floor \cite{Freedman:1973yd, Freedman:1977xn} depicted by the blue dashed curve, which is consistent with the null detection of DM so far in direct detection experiments. On the other hand, for the WIMP-type DM scenario, we can see that some points are already explored by the LUX-ZEPLIN data \cite{LZ:2024zvo}.

\begin{figure}[t!]
	\centering
	\includegraphics[angle=0,height=7cm,width=7.5cm]{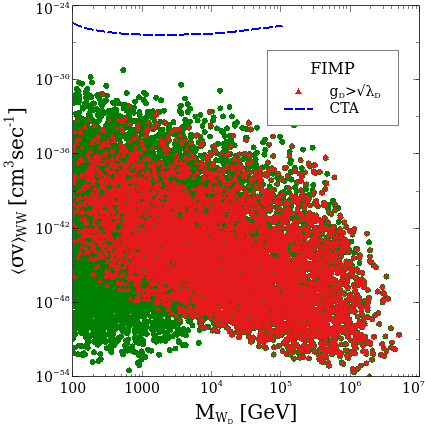}
	\includegraphics[angle=0,height=7cm,width=7.5cm]{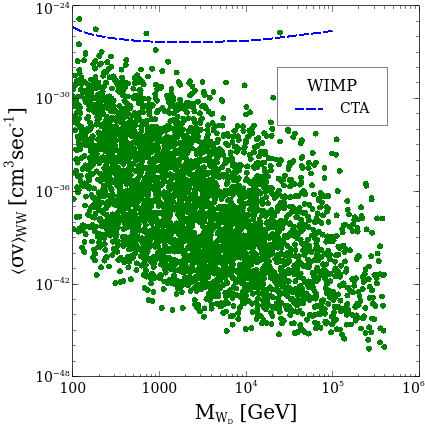}
	\caption{DM indirect detection cross-section times velocity $\langle \sigma v \rangle_{WW}$ in terms of the DM mass $M_{W_D}$. The FIMP-type (WIMP-type) DM scenario is presented in the left (right) panel. The blue dashed curve indicates the projected CTA bound. The colour schemes are the same as those presented in Fig.~\ref{fig:scatter-plot-5}.} 
	\label{fig:scatter-plot-6}
\end{figure}

In Fig.~\ref{fig:scatter-plot-6}, the DM indirect detection cross-section times velocity $\langle \sigma v \rangle_{W W}$, Eq.~\eqref{eqn:ID-cs}, is presented in terms of the DM mass $M_{W_D}$ for the FIMP-type DM (left) and the WIMP-type DM (right). The red points in the left panel satisfy $g_{D} > \sqrt{\lambda_D}$, kinematically opening up the direct decay of the inflaton to DPDM. We observe that $\langle \sigma v \rangle_{W W}$ decreases as the DM mass $M_{W_D}$ increases because of the $1/M^2_{W_D}$ dependence in the cross-section; see Eq.~\eqref{eqn:ID-cs}. As can clearly be seen in the left panel, when DPDM is produced by the freeze-in mechanism, the indirect detection cross-section lies below the future CTA projection \cite{CTA:2020qlo}. However, when DM is produced by the freeze-out mechanism, we see from the right panel that it can be explored in the future run of CTA.

\begin{figure}[t!]
	\centering
	\includegraphics[angle=0,height=7cm,width=7.5cm]{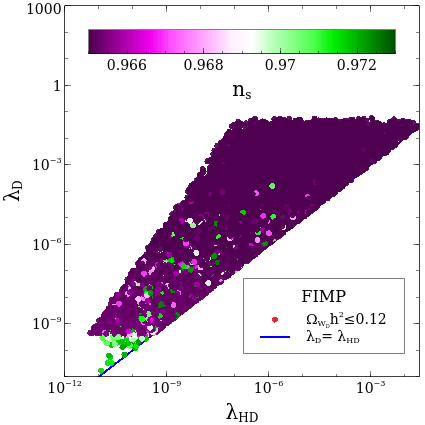}
	\includegraphics[angle=0,height=7cm,width=7.5cm]{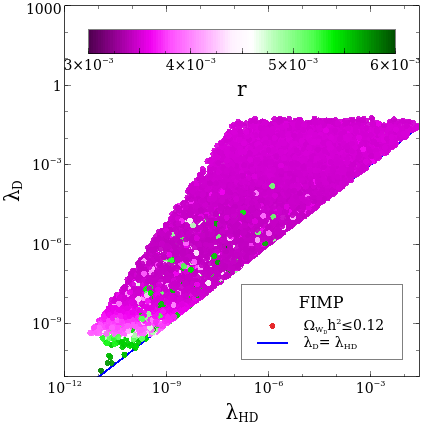}\\
	\includegraphics[angle=0,height=7cm,width=7.5cm]{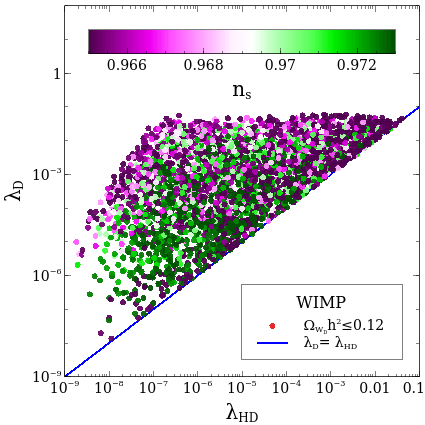}
	\includegraphics[angle=0,height=7cm,width=7.5cm]{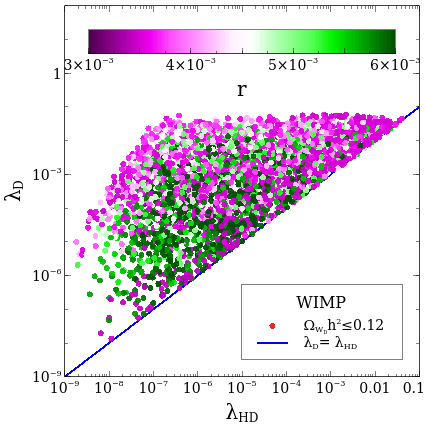}
	\caption{Scalar spectral index $n_s$ and the tensor-to-scalar ratio $r$ in the $\lambda_{HD}$--$\lambda_{D}$ plane. The upper panel corresponds to the FIMP-type DM, while the lower panel is for the WIMP-type DM. The blue line indicates $\lambda_{D} = \lambda_{HD}$. All the points satisfy the relic constraint, $\Omega_{W_D}h^2 \leq 0.12$. We observe that the predicted values for $n_s$ and $r$ lie in the observationally-allowed region, $0.958 \leq n_s \leq 0.975$ and $r \leq 0.036$ \cite{Planck:2018jri, BICEP:2021xfz}. The values of $n_s$ can fall into the latest observational bound reported by the ACT collaboration \cite{ACT:2025tim}; see also Fig.~\ref{fig:ns-r-xiD-plot}.}
	\label{fig:ns-r-plot}
\end{figure}

The predicted values for the spectral index $n_s$ and the tensor-to-scalar ratio $r$ are presented in Fig.~\ref{fig:ns-r-plot}. The upper panel corresponds to the result for the FIMP-type DM, and the lower panel corresponds to the result for the WIMP-type DM. The left panel shows the predicted values for the spectral index $n_s$, while the right panel shows the tensor-to-scalar ratio $r$. After imposing the DM relic constraint, $\Omega_{W_D}h^2 \leq 0.12$, we see that many parameter points survive. We also observe that the resultant values of $n_s$ and $r$ fall into the observationally-favoured region, $0.958 \leq n_s \leq 0.975$ and $r \leq 0.036$ \cite{Planck:2018jri, BICEP:2021xfz}; we also observe that the values of $n_s$ can become compatible with the latest observational bound reported by the ACT collaboration \cite{ACT:2025tim}.

\begin{figure}[t!]
	\centering
	\includegraphics[scale=0.29]{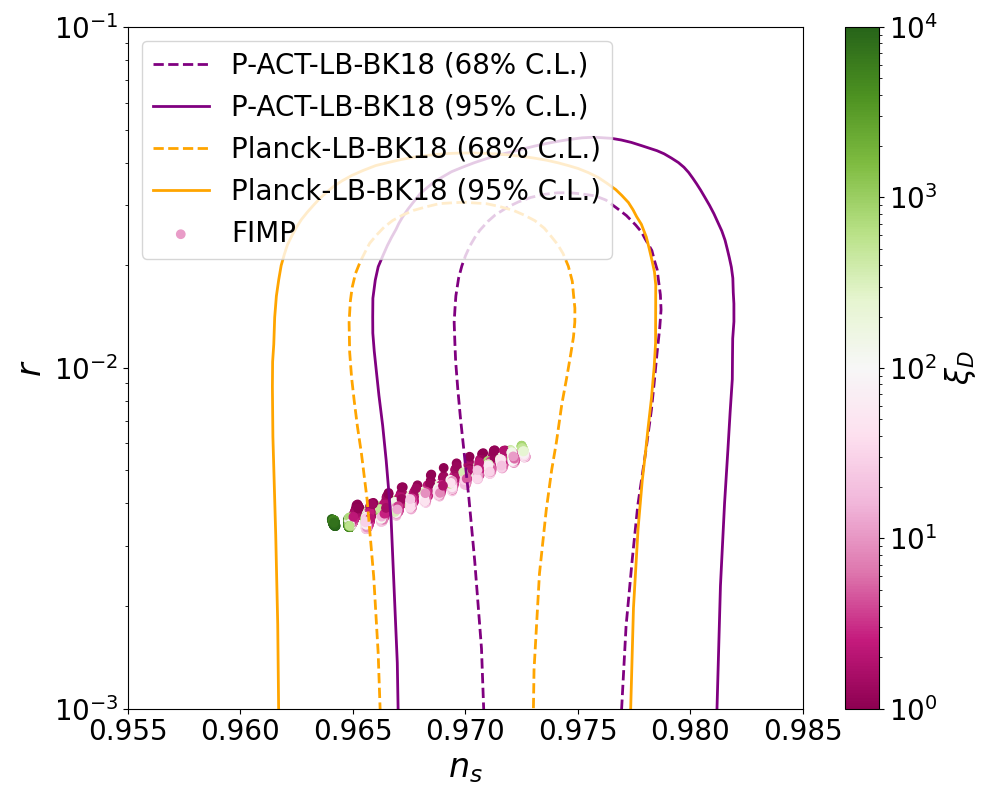}
	\includegraphics[scale=0.29]{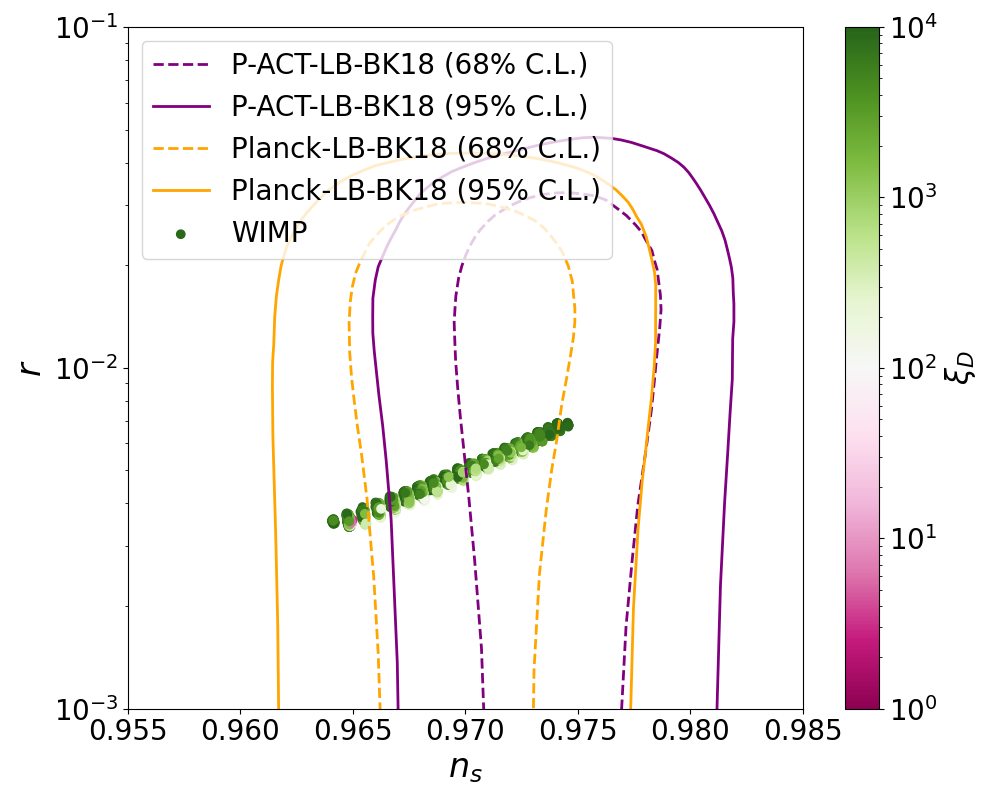}
	\caption{Spectral index $n_s$ and tensor-to-scalar ratio $r$ for the FIMP-type (left) and WIMP-type (right) DM candidates. The points satisfy the DM relic constraint $\Omega_{W_D} h^2 \leq 0.12$. The colour bar represents the value of the nonminimal coupling $\xi_D$. The orange solid (dashed) line is the Planck-BICEP/Keck bound at 95\% (68\%) C.L. \cite{Planck:2018jri, BICEP:2021xfz, ACT:2025tim}, while the purple solid (dashed) line is the latest observational bound coming from the combined analysis of Planck-BICEP/Keck and ACT at 95\% (68\%) C.L. \cite{ACT:2025tim}.} 
	\label{fig:ns-r-xiD-plot}
\end{figure}

The allowed data points in the $n_s$--$r$ plane are shown in Fig.~\ref{fig:ns-r-xiD-plot} for the FIMP-type DM (left panel) and the WIMP-type DM (right panel). The colour represents the value of the nonminimal coupling $\xi_D$. As we have discussed earlier in section~\ref{sec:inflation}, the nonminimal coupling may become as small as of the order unity. The orange lines show the Planck-BICEP/Keck bound \cite{Planck:2018jri, BICEP:2021xfz, ACT:2025tim}, while the purple lines indicate the latest observational bound coming from the combined analysis of Planck-BICEP/Keck and ACT \cite{ACT:2025tim}.\footnote{
	Other recent studies on the compatibility of the Higgs inflationary model and the ACT results include, {\it e.g.}, Refs.~\cite{Gialamas:2025kef, Kim:2025dyi, Liu:2025qca, Zharov:2025evb, Yin:2025rrs, McDonald:2025odl, McDonald:2025tfp}.
}
We see that the predictions of the spectral index $n_s$ and the tensor-to-scalar ratio $r$ are well within the observationally-favoured region.

The model under consideration, namely the dark $U(1)_D$ extension of the SM, can thus incorporate not only both the FIMP-type and the WIMP-type DM candidates, but also inflation with the dark Higgs field as the inflaton. The model studied in the current work naturally aligns well with the current observational constraints, and it has rich prospects to be explored in the near future.

\section{Conclusion}
\label{sec:conc}
In this paper, we have considered a dark $U(1)_D$ gauge extension of the SM and studied the DPDM phenomenology as well as inflationary physics driven by the dark Higgs field nonminimally coupled to gravity. The gauge boson associated with the dark Abelian gauge symmetry, namely the dark photon, becomes a suitable DM candidate. We have considered a low-reheating scenario where DM is produced between the end of inflation and the reheating temperature. Therefore, the produced DM abundance gets diluted due to the entropy production during reheating. With the consideration of such a low reheating temperature, we have achieved both the WIMP- and FIMP-type DM scenarios in the present setup, depending on the strength of the dark gauge coupling. Due to the dilution of DM abundance from the entropy production, we can produce freeze-in DM at a stronger coupling and freeze-out DM at a smaller coupling. As a result, the low-reheating scenario opens up a large parameter space that is yet to be explored in ongoing and future experiments. To check the freeze-out and freeze-in mechanisms of the DM production, we have utilised the inbuilt {\tt micrOMEGAs} routine, which provides information about the nature of the DM production mechanism. The dark gauge coupling mainly determines the DM type, FIMP or WIMP, and is partially influenced by the value of the reheating temperature. We have performed a thorough numerical analysis and illustrated how the DM production responds to the model parameter variations. In particular, we have found that the reheating temperature and the dark gauge coupling play crucial roles. We have also shown that the DM prediction at the direct detection is below the current direct detection limit of LUX-ZEPLIN for FIMP DM and is partly explored for the WIMP DM. Additionally, for indirect detection experiments, the detection prospects of FIMP-type DM will be an order of magnitude below the sensitivity of the proposed CTA experiment, whereas the WIMP-type DM has the possibility to be partly explored.

In addition to DM physics, we have also studied a concrete realisation of cosmic inflation by utilising the dark $U(1)_D$ Higgs field as the inflaton, which nonminimally couples to the Ricci scalar. Taking into account quantum corrections and the running of the couplings, we explicitly demonstrated the predictions on the scalar spectral index $n_s$ and the tensor-to-scalar ratio $r$, and we have identified the allowed parameter space that is favoured by the Planck, BICEP/Keck, and ACT observational data. Moreover, the possibility of achieving a low reheating temperature has been explored in detail. In particular, considering both the decay and scattering processes of the dark Higgs inflaton to the SM Higgs, we have shown that a low reheating temperature could be obtained, depending on the choices of the Higgs quartic couplings. In the case of the decay process, small values of the coupling $\lambda_{HD}$ are preferred to achieve a low reheating temperature. In other words, we need a very small Higgs mixing angle which makes the low reheating temperature scenario less motivated due to slim detection prospects like the standard FIMP-type DM. In the case of the scattering process, if reheating ends when the inflaton field is close to its potential minimum, the possibility of realising a very low reheating temperature with a larger value of $\lambda_{HD}$ or, equivalently, a larger value of the Higgs mixing angle opens up. This latter scenario is attractive as we may have FIMP DM in the detectable regime in current and future experiments.

The current study not only offers a new way of realising the dark gauge boson as a WIMP-type or FIMP-type DPDM candidate with rich detection prospects in ongoing and near-future experiments, but it also provides a way to achieve successful inflation through the dark Higgs field as well as links between DM phenomenology and inflationary physics. In particular, the detailed and concrete demonstration of achieving a low reheating temperature in dark Higgs inflation through the small Higgs portal coupling, $\lambda_{HD} < 10^{-4}$, has important implications for the DM production mechanism with a low reheating temperature.

We note that the reheating dynamics in the dark Higgs inflationary scenario is a three-step process. During the course of reheating, the inflaton potential changes its form twice, from the quadratic one to the quartic form and back to the quadratic form. The first stage of reheating is estimated to be rather short, so one may safely ignore such a period. However, the later two stages need to be properly taken into account. While such a transition is carefully considered and analysed in our inflation discussion, for the DM study, we have assumed that reheating is dominated by the last quadratic regime. This choice is made due to the fact that such a multi-stage reheating scenario is yet to be implemented in {\tt micrOMEGAs}, which supports the matter-like equation of state for the inflaton field. We aim to pursue a dedicated study in the future to properly implement this multi-stage reheating scenario in the DM analysis. Although no qualitative difference from our conclusion is expected, such a study would provide a more complete and consistent framework.

\acknowledgments

S.K. is supported by Brain Pool program funded by the Ministry of Science and ICT through the National Research Foundation of Korea (RS-2024-00407977) and Basic Science Research Program through the National Research Foundation of Korea (NRF) funded by the Ministry of Education, Science and Technology (NRF-2022R1A2C2003567). J.K. is supported by National Natural Science Foundation of China (NSFC) under Grant No. 12505079 and Grant No. 12475060, and by Shanghai Pujiang Program 24PJA134. P.K. is supported by KIAS Individual Grant No. PG021403, and by the Ministry of Science and ICT through the National Research Foundation of Korea (RS-2025-24803289). For the numerical analysis, we have used the Scientific Compute Cluster at GWDG, the joint data centre of Max Planck Society for the Advancement of Science (MPG) and University of G\"{o}ttingen.

\appendix
\section{Expressions for cross-sections and decay width}
\label{apdx:sigmaGamma}
We summarise expressions for the relevant cross-sections and decay width; see also Appendix A of Ref.~\cite{Khan:2023uii}.

\begin{itemize}
\item $\boldsymbol{W_{D}W_{D} \rightarrow W^{+}W^{-}:}$

The cross-section takes 
\begin{align}
    \sigma = \frac{1}{16 \pi s} \left( \frac{s-4M^2_{W}}{s-4M^2_{W_D}} \right)^{1/2}\, |M|^2_{WW}
    \,,
\end{align}
where $s$ is the Mandelstam variable, and $|M|_{WW}$ is the amplitude given by
\begin{align}
    |M|^{2}_{WW} &= \frac{4}{9}  \left| \frac{g_{h_{1}W_{D}W_{D}} g_{h_{1}WW }}{(s-M^2_{h_1}) 
    + i \Gamma_{h_1} M_{h_1}} + \frac{g_{h_{2}W_{D}W_{D}} g_{h_{2}WW }}{(s-M^2_{h_2}) 
    + i \Gamma_{h_2} M_{h_2}} \right|^{2} 
    \nonumber \\
    &\quad
    \times \left( 1 + \frac{(s-2M^2_{W_{D}})^2}{8 M^4_{W_{D}}}\right) \left( 1 + \frac{(s-2M^2_{W})^2}{8 M^4_{W}}\right)\,.
\end{align}
The vertices are
\begin{align}
    g_{h_{1(2)W_{D}W_{D}}} &= -2 g_{D} M_{W_{D}} \sin \alpha (-\cos\alpha)
    \,,\nonumber \\
    g_{h_{1(2)WW}} &= \frac{v}{2 s^2_w} \cos \alpha\, (\sin\alpha)
    \,,
    \label{eqn:vertex-expre}
\end{align}
where $s^2_w = 0.23$.

\item $\boldsymbol{W_{D}W_{D} \rightarrow ZZ:}$

The cross-section for this process takes
\begin{align}
    \sigma = \frac{1}{32 \pi s} \left( \frac{s-4M^2_{Z}}{s-4M^2_{W_D}} \right)^{1/2}\, |M|^2_{ZZ}\,,
\end{align}
where the amplitude $|M|_{ZZ}$ is given by
\begin{align}
    |M|^{2}_{ZZ} &= \frac{4}{9}  \left| \frac{g_{h_{1}W_{D}W_{D}} g_{h_{1}ZZ }}{(s-M^2_{h_1}) 
    + i \Gamma_{h_1} M_{h_1}} + \frac{g_{h_{2}W_{D}W_{D}} g_{h_{2}ZZ }}{(s-M^2_{h_2}) 
    + i \Gamma_{h_2} M_{h_2}} \right|^{2} 
    \nonumber \\
    &\quad 
    \times \left( 1 + \frac{(s-2M^2_{W_{D}})^2}{8 M^4_{W_{D}}}\right) \left( 1 + \frac{(s-2M^2_{Z})^2}{8 M^4_{Z}}\right)\,,
\end{align}
with the vertex
\begin{align}
    g_{h_{1(2)ZZ}} = \frac{v}{2 c^2_w s^2_w} \cos \alpha \, (\sin\alpha)\,.
\end{align}
Here, $c^2_{w} = 1 - s^2_{w}$.

\item $\boldsymbol{W_{D}W_{D} \rightarrow f \bar{f}:}$

The cross-section for this process takes
\begin{align}
    \sigma = \frac{1}{16 \pi s} \left( \frac{s-4M^2_{f}}{s-4M^2_{W_D}} \right)^{1/2}\, |M|^2_{ff}
    \,,
\end{align}
where the amplitude $|M|_{ff}$ is given by
\begin{align}
    |M|^{2}_{ff} &= \frac{4}{9} \left(s - 4 M^2_{f} \right)  \left| \frac{g_{h_{1}W_{D}W_{D}} g_{h_{1}ff }}{(s-M^2_{h_1}) 
    + i \Gamma_{h_1} M_{h_1}} + \frac{g_{h_{2}W_{D}W_{D}} g_{h_{2}ff }}{(s-M^2_{h_2}) 
    + i \Gamma_{h_2} M_{h_2}} \right|^{2} 
    \nonumber \\
    &\quad 
    \times \left( 1 + \frac{(s-2M^2_{W_{D}})^2}{8 M^4_{W_{D}}}\right)\,,
\end{align}
with
\begin{align}
    g_{h_{1(2)ff}} &= - \frac{M_{f}}{v} \cos \alpha\, (\sin\alpha)\,.
\end{align}

\item $\boldsymbol{W_{D}W_{D} \rightarrow h_{i}h_{j}:}$

The cross-section for this process takes
\begin{align}
    \sigma = \frac{1}{16 \pi s S_{ij}} \left( \frac{s-4M^2_{f}}{s-4M^2_{W_D}} \right)^{1/2}\, |M|^2_{h_{i}h_{j}}
    \,,
\end{align}
where $S_{ij} = 1 (2)$ for $i\neq j$ $(i=j)$ with $i,j = 1,2$, and the amplitude $|M|_{h_{i}h_{j}}$ is given by
\begin{align}
    |M|^{2}_{h_{i}h_{j}} &=
    \frac{2}{9} \left| \frac{g_{h_{1}W_{D}W_{D}} g_{h_{1}h_{i}h_{j} }}{(s-M^2_{h_1}) 
    + i \Gamma_{h_1} M_{h_1}} + \frac{g_{h_{2}W_{D}W_{D}} g_{h_{2}h_{i}h_{j} }}{(s-M^2_{h_2}) 
    + i \Gamma_{h_2} M_{h_2}} - g_{W_{D}W_{D}h_{i}h_{j}}\right|^{2} 
    \nonumber \\
    &\quad 
    \times \left( 1 + \frac{(s-2M^2_{W_{D}})^2}{8 M^4_{W_{D}}}\right)
    \,,
\end{align}
with the vertices
\begin{align}
    g_{h_{2}h_{2}h_{2}} &=  
    -3 \bigl[ \lambda_{HD} \sin\theta \cos\theta (v_{D} \sin\theta + v \cos\theta) + 2 \lambda_{D}  v_{D} \cos^{3}\theta + 2 \lambda_{H} v \sin^{3}\theta 
    \bigr]
    \,,\nonumber \\
    g_{h_{1}h_{1}h_{1}} &=  
    3 \bigl[ \lambda_{HD} \sin\theta \cos\theta (v_{D} \cos\theta - v \sin\theta) + 2 \lambda_{D} v_{D} \sin^{3}\theta - 2 \lambda_{H} v \cos^{3}\theta 
    \bigr]
    \,,\nonumber \\
    g_{h_{1}h_{2}h_{2}} &=  
    2 (3 \lambda_{D} - \lambda_{HD}) v_{D} \sin\theta \cos^{2}\theta
    + 2 (-3 \lambda_{H} + \lambda_{HD}) v \cos\theta \sin^{2}\theta 
    \nonumber \\
    &\quad
    + \lambda_{HD} (v_{D} \sin^{3}\theta - v \cos^{3}\theta) 
    \,,\nonumber \\
    g_{h_{2}h_{1}h_{1}} &=  
    2 (-3 \lambda_{H} + \lambda_{HD}) v \sin\theta \cos^{2}\theta
    + 2 (-3 \lambda_{D} + \lambda_{HD}) v_{D} \cos\theta \sin^{2}\theta   
    \nonumber \\
    &\quad   
    - \lambda_{HD} (v_{D} \cos^{3}\theta + v \sin^{3}\theta)
    \,,\nonumber \\
    g_{W_{D}W_{D}h_{2}h_{2}} &= 2 \cos^{2}\theta g^2_{D}
    \,,\nonumber\\
    g_{W_{D}W_{D}h_{1}h_{1}} &= 2 \sin^{2}\theta g^2_{D}
    \,,\nonumber\\
    g_{W_{D}W_{D}h_{1}h_{2}} &=  -2\cos\theta \sin\theta g^2_{D} 
    \,.
\end{align}

\item $\boldsymbol{A \rightarrow W_{D}W_{D}:}$

The decay width for this process takes
\begin{align}
    \Gamma_{A \rightarrow W_{D}W_{D}} = \frac{M^3_{A} g^2_{A W_{D}W_{D}}}{128 \pi M^4_{W_D}} \sqrt{1 - \frac{4 M^2_{W_D}}{M^2_{A}}} \left( 1 - 
    \frac{4 M^2_{W_D}}{M^2_{A}} + \frac{12 M^2_{W_D}}{M^2_{A}} \right)
    \,,
\end{align}
where the vertex factor $g_{AW_{D}W_{D}}$ is given in Eq.~\eqref{eqn:vertex-expre} for $A = h_{1,2}$.
\end{itemize}

\section{Renormalisation group equations}
\label{apdx:RGEs}
The beta functions are given by (see also Ref.~\cite{Khan:2023uii})
\begin{align}
	(4\pi)^2\beta_{g_1} &= \frac{81+s_H}{12}g_1^3 \,,\\
	(4\pi)^2\beta_{g_2} &= -\frac{39-s_H}{12}g_2^3 \,,\\
	(4\pi)^2\beta_{g_3} &= -7g_3^3 \,,\\
	(4\pi)^2\beta_{g_D} &= \frac{s_D}{3}g_D^3 \,,\\
	(4\pi)^2\beta_{y_t} &= y_t\left[
	\left(\frac{23}{6}+\frac{2}{3}s_H\right)y_t^2
	-\left(
	8g_3^2 + \frac{17}{12} g_1^2 + \frac{9}{4}g_2^2
	\right)
	\right]
	\,,\\
	(4\pi)^2\beta_{\lambda_H} &=
	6(1+3s_H^2)\lambda_H^2 + \frac{1+s_D^2}{2}\lambda_{HD}^2 - 3g_1^2\lambda_H
	-9g_2^2\lambda_H 
	\nonumber\\
	&\quad
	+ \frac{3}{8}g_1^4 + \frac{3}{4}g_1^2g_2^2
	+\frac{9}{8}g_2^4 + 12\lambda_H y_t^2 - 6y_t^4
	\,,\\
	(4\pi)^2\beta_{\lambda_D} &=
	2(1+9s_D^2)\lambda_D^2 + \frac{3+s_H^2}{2}\lambda_{HD}^2
	- 12g_D^2\lambda_D + 6g_D^4
	\,,\\
	(4\pi)^2\beta_{\lambda_{HD}} &=
	6(1+s_H^2)\lambda_H\lambda_{HD} + 2(1+3s_D^2)\lambda_D\lambda_{HD}
	+4s_Hs_D\lambda_{HD}^2
	\nonumber\\
	&\quad
	- \frac{3}{2}g_1^2\lambda_{HD}
	- \frac{9}{2}g_2^2\lambda_{HD} 
	- 6g_D^2\lambda_{HD}
	+ 6\lambda_{HD}y_t^2
	\,,\\
	(4\pi)^2\beta_{\xi_H} &=
	\left[
	6(1+s_H)\lambda_H - \frac{3}{2}(g_1^2+3g_2^2) + 6y_t^2
	\right]\left(
	\xi_H + \frac{1}{6}
	\right)
	\nonumber\\&\quad
	+(1+s_D)\lambda_{HD}\left(
	\xi_D + \frac{1}{6}
	\right)
	\,,\\
	(4\pi)^2\beta_{\xi_D} &=
	\left[
	2(1+3s_D)\lambda_D - 6g_D^2
	\right]\left(
	\xi_D + \frac{1}{6}
	\right)
	+ (3+s_H)\lambda_{HD}\left(
	\xi_H + \frac{1}{6}
	\right)
	\,,
\end{align}
with the suppression factors
\begin{align}
	s_H = \frac{1+\xi_H h^2/M_{\rm P}^2}{1+(1+6\xi_H)\xi_H  h^2/M_{\rm P}^2}
	\,,\quad
	s_D = \frac{1+\xi_D \phi^2/M_{\rm P}^2}{1+(1+6\xi_D)\xi_D  \phi^2/M_{\rm P}^2}
	\,.
\end{align}


\bibliographystyle{JHEP}
\bibliography{main}
\end{document}